\documentclass[preprint,3p,12pt]{elsarticle}
\usepackage{mathrsfs}
\usepackage{amsmath}
\usepackage{stmaryrd}
\usepackage{bbding}
\usepackage{dcolumn}
\usepackage{amsfonts}
\usepackage{amssymb}
\usepackage{psfrag}
\usepackage{wrapfig}
\usepackage{makeidx}
\usepackage{bm}
\usepackage{epsf}
\usepackage{epsfig}
\usepackage{setspace}
\usepackage{epstopdf}
\usepackage{psfrag}
\usepackage{color}
\usepackage{scalerel}  
\usepackage{lineno}
\usepackage{graphicx}
\usepackage{caption}
\usepackage{subcaption}
\usepackage{overpic}
\usepackage{hyperref} 
\usepackage{verbatim}

\begin{document}
%\title{An efficient and robust high-order compact ALE gas-kinetic scheme for unstructured meshes}
\title{High-order compact ALE gas-kinetic scheme on three-dimensional unstructured meshes}

\author[BNU]{Yibo Wang}
\ead{wangybbnu@163.com}

\author[XJTU]{Xing Ji}
\ead{jixing@xjtu.edu.cn}

\author[BNU]{Liang Pan\corref{cor}}
\ead{panliang@bnu.edu.cn}
\cortext[cor]{Corresponding author}

\address[BNU]{Laboratory of Mathematics and Complex Systems, School of Mathematical Sciences, Beijing Normal University, Beijing, China}
 
\address[XJTU]{State Key Laboratory for Strength and Vibration of Mechanical Structures, School of Aerospace Engineering, Xi’an Jiaotong University, Xi’an, China}

\begin{abstract}
In this paper, a compact high-order gas-kinetic scheme (HGKS) is developed for compressible arbitrary Lagrangian-Eulerian (ALE) computations on three-dimensional unstructured meshes. 
To achieve high-order spatial accuracy, a large number of matrix inversions are required for spatial reconstruction. 
In ALE computations, the mesh moves and the geometric information must be recalculated at each time step, which reduces computational efficiency. 
To improve efficiency, a memory-reduction reconstruction strategy is employed to construct a quadratic polynomial on unstructured meshes. 
By differentiating the candidate polynomial, the quadratic terms are obtained by the least-squares method using the average gradient values of the cell itself and its neighboring cells. 
The linear terms of the polynomial are also determined by the least-squares method. 
On three-dimensional unstructured meshes, a $3\times3$ coefficient matrix must be stored, and four least-squares procedures are required to construct a quadratic polynomial. 
To suppress numerical oscillations in flows with discontinuities, generalized essentially non-oscillatory (GENO) reconstruction and a gradient compression factor are adopted. 
To accelerate the computation, a multi-GPU accelerated code is implemented using Compute Unified Device Architecture (CUDA) and Message Passing Interface (MPI). 
The current scheme is fully explicit, and the acceleration framework can be developed based on cell-level parallelism. 
Numerical examples are presented to evaluate the accuracy and robustness of the scheme, including flows with strong discontinuities, pitching wings, and helicopter rotors. 
Numerical results confirm the excellent performance of the compact ALE scheme on unstructured meshes.
\end{abstract}

\begin{keyword}
Gas-kinetic scheme, arbitrary Lagrangian-Eulerian (ALE) method, memory-reduction compact scheme.  
\end{keyword}

\maketitle

\section{Introduction}

Flows with moving boundaries are encountered in a wide range of aerospace applications, 
including flapping wings, pitching wings and rotating helicopter rotors. 
The accurate simulation of these problems requires numerical methods to handle different boundary motions and resolve the unsteady flow structures. 
Various approaches have been developed for these problems, 
such as the overset-grid methods \cite{Overset-Steger}, 
immersed-boundary approaches \cite{IBM-Mittal} including direct-forcing methods \cite{IBM-Uhlmann,IBM-direct-Meng} and cut-cell method \cite{IBM-cutcell}, 
and direct arbitrary-Lagrangian-Eulerian (ALE) method \cite{ALE-direct-1,ALE-direct-2}. 
Among these methods, the direct ALE method adopts a body-fitted mesh that follows the moving body.
The wall boundary conditions can be directly imposed and the near-wall mesh spacing can be controlled to resolve the micro-scale flow structures. 
The ALE method was originally developed by Hirt \cite{ALE-1,ALE-2}, in which the mesh motion can be chosen arbitrarily, providing additional flexibility and accuracy.  
There are two types of ALE methods, i.e. the direct and indirect ALE methods. 
The indirect ALE methods are divided into three distinct stages,
i.e., Lagrangian stage \cite{Lagrangian-Benson} , rezoning stage \cite{ALE-rezone} , and remapping stage \cite{ALE-remap}. 
The direct ALE method is based on the space-time conservation formulation of the governing equations, 
where the rezoned geometry is taken into account during the computation and the remapping stage is omitted. 
The grid velocity $\boldsymbol{U}_g$ can be arbitrarily specified for the control volume. 
The Eulerian and Lagrangian descriptions are recovered by setting $\boldsymbol{U}_g=0$ and $\boldsymbol{U}_g=\boldsymbol{U}$, 
where $\boldsymbol{U}$ denotes the fluid velocity. These features make the direct ALE framework an attractive choice for moving-boundary problems.

In the past decades, the gas-kinetic schemes (GKS) based on the Bhatnagar-Gross-Krook (BGK) model \cite{BGK-1,BGK-2} have been
developed systematically for the computations from low speed flows to supersonic ones \cite{GKS-Xu1,GKS-Xu2}. 
The gas-kinetic scheme presents a gas evolution process from the kinetic scale to hydrodynamic scale, 
and  both inviscid and viscous fluxes can be calculated in one framework. 
The two-stage fourth-order temporal discretization, originally developed for Lax‑Wendroff type flow solvers \cite{GRP-high}, 
provides a reliable framework for constructing gas‑kinetic schemes with fourth‑order temporal accuracy \cite{GKS-high-1,GKS-high-2}.
With the WENO reconstruction \cite{simple-WENO-1,simple-WENO-2}, 
the third-order gas-kinetic schemes on three-dimensional unstructured meshes \cite{GKS-noncompact-1,GKS-noncompact-2} are developed, 
in which a simple strategy of selecting stencils for reconstruction is adopted and the topology independent linear weights are used. 
Based on the unified coordinate transformation, a second-order gas-kinetic scheme was developed under the moving-mesh framework  \cite{GKS-ALE1}. 
With the integral form of the fluid dynamic equations, a second-order remapping-free ALE gas-kinetic scheme was developed \cite{GKS-ALE2},  which can be considered as a direct ALE method. 
The grid velocity can be chosen arbitrarily in the flux calculation, and the rezoning and remapping stages in the indirect ALE method are avoided.
For high-order accuracy, the one-stage DG-ALE gas-kinetic method was developed, especially for the oscillating airfoil calculations  \cite{GKS-ALE3}. 
However, the one-stage gas evolution model and DG framework become very complicated, 
and the efficiency of the DG-ALE-GKS becomes low due to the severely constrained time step from the DG formulation.
With the WENO reconstructions \cite{WENO-unstructured, simple-WENO-1,simple-WENO-2}, 
 the high-order gas-kinetic schemes are also developed for the ALE computations \cite{GKS-ALE4,GKS-ALE5},
in which the expected order of accuracy is achieved and the geometric conservation law is well preserved. 
However, the geometric information needs to be calculated at each time step due to the mesh movement. 
Such calculation includes a large number of matrix inversion operations in the WENO reconstructions, 
which affect the efficiency of computation dramatically. 
Recently, a memory-reduction compact gas-kinetic scheme has been proposed \cite{GKS-MR-compact} , in which a quadratic polynomial can be constructed 
on the target cell efficiently.  For the two-dimensional case, only one $2\times2$ coefficient matrix needs to be stored, and three times of least square
process are needed to construct a quadratic polynomial. With the memory-reduction compact reconstruction, 
an efficient and robust high-order compact gas-kinetic scheme for ALE computation is developed  \cite{GKS-ALE6}.
Compared with the previous high-order ALE method, the efficiency is improved greatly in the process of mesh movement.

In this paper, the high-order gas-kinetic scheme with memory-reduction compact reconstruction will be extended to the three-dimensional ALE computation.
To deal with the complicated geometry, the hybrid unstructured meshes with tetrahedral, pyramidal, prismatic and hexahedral cells are considered.
To achieve the third-order accuracy, a quadratic polynomial needs to be constructed in a cell.  
Taking derivatives of the candidate polynomial, the quadratic terms can be obtained by the least
square method using the average gradient values of the cell itself and its adjacent cells. Moving the quadratic terms to right-hand
side of the constraints for cell averaged value, the linear terms of the polynomial can be determined by the least square method.
For the three-dimensional case, the $3\times3$ coefficient matrix needs to be stored, and four times of least square
process are needed to construct a quadratic polynomial. 
To suppress numerical oscillations in flow with discontinuities, the generalized ENO (GENO) nonlinear reconstruction \cite{GKS-GENO} 
and  the gradient compression factor \cite{GKS-compact-gcf} are  adopted.
To implement large scale computation and accelerate the efficiency, the compact ALE gas-kinetic scheme is also implemented with multiple GPUs 
using Compute Unified Device Architecture (CUDA) and Message Passing Interface (MPI) architecture.
The domain is decomposed by METIS \cite{METIS-Kar}, and the MPI processes are created for the control of each process and communication among GPUs. 
Numerical examples are presented to evaluate the accuracy, efficiency, robustness, and the preservation of geometric conservation law of the current scheme.

This paper is organized as follows. In Section 2, the gas-kinetic scheme in ALE formulation is introduced. 
The memory-reduction compact reconstruction is presented in Section 3. In Section 4, the GPU implementation is presented. 
Section 5 includes numerical examples to validate the current scheme.
The last section is the conclusion.

\section{Gas-kinetic scheme in ALE formulation}
The three-dimensional BGK equation \cite{BGK-1,BGK-2} can be written as
\begin{equation}\label{bgk}
f_t+uf_x+vf_y+wf_z=\frac{g-f}{\tau},
\end{equation}
where $\boldsymbol{u}=(u,v,w)$ is the particle velocity, $f$ is the
gas distribution function, $g$ is the three-dimensional Maxwellian
distribution and $\tau$ is the collision time. The collision term
satisfies the compatibility condition
\begin{equation}\label{compatibility}
\int \frac{g-f}{\tau}\psi \text{d}\Xi=0,
\end{equation}
where
$\displaystyle\psi=(\psi_1,...,\psi_5)^T=(1,u,v,w,\frac{1}{2}(u^2+v^2+w^2+\xi^2))^T$,
the internal variable $\xi^2=\xi^2_1+…+\xi^2_K$,
$\text{d}\Xi=\text{d}u\text{d}v\text{d}w\text{d}\xi^1...\text{d}\xi^{K}$,
$\gamma$ is the specific heat ratio and  $K=(5-3\gamma)/(\gamma-1)$
is the degree of freedom for three-dimensional flows.

According to the Chapman-Enskog expansion for BGK equation, the macroscopic
governing equations can be derived \cite{GKS-Xu1,GKS-Xu2}. In the
continuum region, the BGK equation can be rearranged and the gas
distribution function can be expanded as
\begin{align*}
f=g-\tau D_{\boldsymbol{u}}g+\tau D_{\boldsymbol{u}}(\tau
D_{\boldsymbol{u}})g-\tau D_{\boldsymbol{u}}[\tau
D_{\boldsymbol{u}}(\tau D_{\boldsymbol{u}})g]+...,
\end{align*}
where $D_{\boldsymbol{u}}=\displaystyle\frac{\partial}{\partial
t}+\boldsymbol{u}\cdot \nabla$. With the zeroth-order truncation
$f=g$, the Euler equations can be obtained. For the first-order
truncation
\begin{align*}
f=g-\tau (ug_x+vg_y+wg_z+g_t),
\end{align*}
the Navier-Stokes equations can be obtained. With higher-order
truncations, the Burnett and super-Burnett equations can be obtained
as well.

To develop the numerical scheme in the arbitrary-Lagrangian-Eulerian (ALE) framework, the BGK equation on moving reference can be considered
\begin{equation}\label{bgk-moving}
f_t+(u-U^g)f_x+(v-V^g)f_y+(w-W^g)f_z =\frac{g-f}{\tau},
\end{equation}
where $\boldsymbol{U}^{g}=(U^g,V^g,W^g)$ is the grid velocity given arbitrarily. Taking moments of Eq.\eqref{bgk-moving}
and integrating with respect to space, the semi-discretized finite
volume scheme can be expressed as
\begin{align}\label{semi}
\frac{\text{d}(|\Omega_{i}|Q_{i})}{\text{d}t}=-\mathcal{L}(\Omega_{i},Q_{i}),
\end{align}
where $Q_{i}$ is the cell averaged conservative value of
$\Omega_{i}$ and $|\Omega_{i}|$ is the volume of $\Omega_{i}$. The operator $\mathcal{L}$ is defined as
\begin{equation}\label{finite}
\mathcal{L}(\Omega_{i},Q_{i})=-\sum_{p \in N(i)}\mathbb{F}_{p}(t)=-\sum_{p \in N(i)}\iint_{\Sigma_{p}}\boldsymbol{F}(Q,t)\text{d}\sigma,
\end{equation}
where $N(i)$ denotes the interfaces surrounding cell $i$, $\Sigma_{p}$ is the $pth$ cell interface of $\Omega_{i}$, 
$\boldsymbol{n}$ is the outer normal direction and the numerical flux
is given as
\begin{align*}
\boldsymbol{F}(Q,t)=\int\psi f(\boldsymbol{x},t,\boldsymbol{u},\xi)
(\boldsymbol{u}-\boldsymbol{U}^g) \cdot
\boldsymbol{n}\text{d}\Xi.
\end{align*}
%The structured mesh is used in this paper, and the extension to unstructured mesh will be considered in the future.

Based on the time-dependent flux function of the generalized Riemann problem solver (GRP) 
\cite{GRP-high} and gas-kinetic scheme \cite{GKS-Xu1,GKS-Xu2}, 
a two-stage fourth-order time-accurate discretization was developed for Lax-Wendroff type flow solvers. 
In this study, the two-stage method is used for gas-kinetic solvers. 
Consider the following time-dependent equation
\begin{align*}
\frac{\partial Q}{\partial t}=\mathcal {L}(Q),
\end{align*}
where $\mathcal {L}$ is an operator for spatial derivative of flux, 
the flow variable $Q^{n+1}$ at $t_{n+1}=t_n+\Delta t$ can be updated with the following formula
\begin{align*}
&Q^*    = Q^n+\frac{1}{2}\Delta t\mathcal{L}(Q^n)+\frac{1}{8}\Delta t^2\partial_t\mathcal{L}(Q^n), \\
 Q^{n+1}=&Q^n+\Delta t\mathcal {L}(Q^n)
        +\frac{1}{6}\Delta t^2\big(\partial_t\mathcal{L}(Q^n)+2 \partial_t\mathcal{L}(Q^*)\big),
\end{align*}
where $Q^*$ is the variables at $t^*=t_n+\Delta t/2$.
It can be proved that for hyperbolic equations the above temporal discretization provides a fourth-order time accurate solution. 
To implement the two-stage fourth-order method for Eq.\eqref{finite}, 
$\mathcal{L}$ and $\partial_t\mathcal{L}$ at the initial and intermediate stages need to be given. 
For the moving-mesh computation, the temporal discretization above can be extended to Eq.\eqref{finite} by denoting 
\begin{align*}
\frac{\text{d}(|\Omega_i|Q_{i})}{\text{d}t}=\mathcal{L}(\Omega_{i},Q_{i})\triangleq-\sum_{p \in N(i)}\mathbb{F}_{p}(t).
\end{align*}
The two-stage fourth-order time-accurate discretization is rewritten as 
\begin{align*}
(|\Omega^*|Q^*) &= (|\Omega^n|Q^n)+\frac{1}{2}\Delta t\mathcal{L}(\Omega^n,Q^n)+\frac{1}{8}\Delta t^2\partial_t\mathcal{L}(\Omega^n,Q^n), \\
(|\Omega^{n+1}|Q^{n+1})=&(|\Omega^n|Q^n)+\Delta t\mathcal {L}(\Omega^n,Q^n)
        +\frac{1}{6}\Delta t^2\big(\partial_t\mathcal{L}(\Omega^n,Q^n)+2\partial_t\mathcal{L}(\Omega^*,Q^*)\big),
\end{align*}
where $\mathcal{L}$ and $\partial_t\mathcal{L}$ at $t^n$ can be given by
\begin{equation}\label{two-stage-2}
\begin{aligned}
\mathcal{L}(\Omega_{i}^n,Q_{i}^n)=&-\sum_{p \in N(i)}\mathbb{F}_{p}^n, \\
\partial_t\mathcal{L}(\Omega_{i}^n,Q_{i}^n)=&-\sum_{p \in N(i)}\partial_t\mathbb{F}_{p}^n,
\end{aligned}
\end{equation}
and $\partial_t\mathcal{L}$ at  $t^*=t_n+\Delta t/2$ can be given similarly.
To determine these coefficients, the time dependent numerical flux $\mathbb{F}_{p}(t)$ in Eq.\eqref{two-stage-2} can be approximated as a linear function
\begin{align}\label{expansion-1}
\mathbb{F}_{p}(t)=\mathbb{F}_{p}^n+ \partial_t\mathbb{F}_{p}^n(t-t_n).
\end{align}
For the gas-kinetic scheme, the gas evolution is a relaxation process from kinetic to hydrodynamic scale through the exponential function, and the corresponding flux is a complicated function of time. 
In order to obtain the time derivatives of the flux function, the flux function should be approximated as a linear function of time within a time interval. 
Integrating Eq.\eqref{expansion-1} over $[t_n, t_n+\Delta t/2]$ and
$[t_n, t_n+\Delta t]$, we have the following two equations
\begin{align}\label{expansion-2}
\begin{cases}
\displaystyle\mathbb{F}_{p}^n\Delta t+\frac{1}{2}\partial_t\mathbb{F}_{p}^n\Delta t^2 =\int_{t_n}^{t+_n\Delta t}\mathbb{F}_{p}(t)  \text{d}t, \\
\displaystyle\frac{1}{2}\mathbb{F}_{p}^n\Delta t+\frac{1}{8}\partial_t\mathbb{F}_{p}^n\Delta t^2 =\int_{t_n}^{t_n+ \Delta t/2}\mathbb{F}_{p}(t)\text{d}t.
\end{cases}
\end{align}
The coefficients $\mathbb{F}_{p}^n$ and $\partial_t\mathbb{F}_{p}^n$
can be obtained by solving the linear system, and
$\mathcal{L}(\Omega_{i}^n,Q_{i}^n)$ and
$\partial_t\mathcal{L}(\Omega_{i}^n,Q_{i}^n)$ can be given by
Eq.\eqref{two-stage-2}.

For the moving-mesh computation, the variation of geometric variable and grid velocity need to be considered
for the numerical flux to preserve the geometric conservation law (GCL). 
Generally, the GCL is referred to as the preservation of uniform flow with arbitrarily 
moving grids for any numerical discretization scheme \cite{GCL-1,GCL-2}. For the current scheme, 
the technique for preservation of GCL in \cite{GKS-ALE5} is adopted.

\begin{figure}[!h]
\centering
\includegraphics[width=0.55\textwidth]{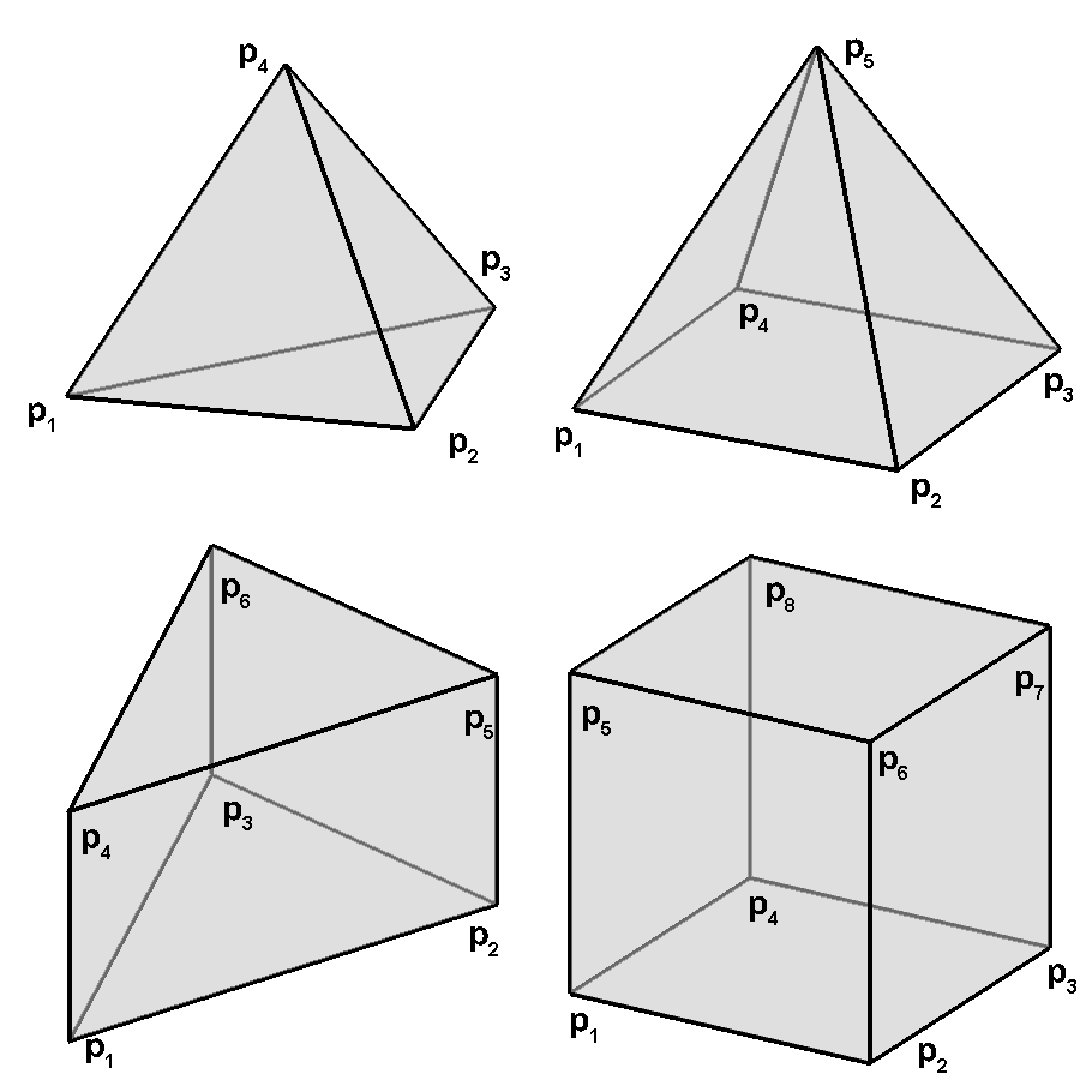}
\caption{\label{schematic-cell} Schematic for tetrahedron, pyramid, prism and hexahedron.}
\end{figure}

\section{Memory-reduction compact reconstruction} 
For the unstructured hybrid meshes, the meshes consist of the
tetrahedral, pyramidal, prismatic and hexahedral cells. The
schematic of the cells are given in Fig.\ref{schematic-cell} with
the label of vertexes. For the sake of clearness, the faces of
control volume also need to be labeled. For the cell $\Omega_0$, the
faces are labeled as follows
\begin{itemize}
\item For the tetrahedral cell, four faces are denoted as
\begin{align*}
F_{1}=\{p_1p_2p_3\}, F_{2}=\{p_1p_2p_4\},
F_{3}=\{p_2p_3p_4\}, F_{4}=\{p_3p_1p_4\}.
\end{align*}
\item For the pyramidal cell, five faces are denoted as
\begin{align*}
F_{1}&=\{p_1p_2p_5\}, F_{2}=\{p_2p_3p_5\},F_{3}=\{p_3p_4p_5\}, \\
F_{4}&=\{p_4p_1p_5\}, F_{5}=\{p_1p_2p_3p_4\}.
\end{align*}
\item For the prismatic cell, five faces are denoted as
\begin{align*}
F_{1}&=\{p_1p_2p_3\}, F_{2}=\{p_4p_5p_6\},\\
F_{3}&=\{p_1p_2p_6p_4\}, F_{4}=\{p_2p_3p_6p_5\}, F_{5}=\{p_3p_1p_4p_6\}.
\end{align*}
\item For the hexahedral cell, six faces are denoted as
\begin{align*}
F_{1}&=\{p_1p_2p_3p_4\}, F_{2}=\{p_1p_2p_6p_5\}, F_{3}=\{p_2p_3p_7p_6\},\\
F_{4}&=\{p_3p_4p_8p_7\}, F_{5}=\{p_4p_1p_5p_8\}, F_{6}=\{p_5p_6p_7p_8\}.
\end{align*}
\end{itemize} 
With the labeled faces given above, the neighboring cell of
$\Omega_{i}$, which shares the face $F_{p}$, is denoted as
$\Omega_{i_p}$. 
If $\Omega_{i}$ is a boundary cell,  the neighboring cells of
$\Omega_{i}$ can be found with the periodic boundary condition.
Thus,  for each cell $\Omega_{i}$ in the domain, the big stencil
$S_i$ can be selected.  
If $\Omega_{i}$ is a boundary cell, the
geometric information and flow variables of the ghost cell
$\Omega_{i_{p}}$ are provided  by the mirror image of
neighboring cells of $\Omega_{i}$.

\subsection{High-order spatial reconstruction}
Different from the traditional Riemann solvers, the gas-kinetic scheme provides a time-dependent gas distribution function. 
Meanwhile, the macroscopic conservative variables can be obtained by taking moments of the distribution function as well 
\begin{align}\label{point}
Q(\boldsymbol{x}_{p,m},t)=\int\boldsymbol\psi f(\boldsymbol{x}_{p,m},t,\boldsymbol{u},\xi)\text{d}\Xi, 
\end{align}
where $f(\boldsymbol{x}_{p,m},t,\boldsymbol{u},\xi)$ is the gas distribution function at $m$-th Gaussian point on the  interface $p$. 
According to the Gauss's theorem, the cell-averaged gradient of the flow variable $Q$ can be calculated as follows 
\begin{equation*}
\begin{split}
|\Omega_i|(\nabla Q)_i(t)&=\iint_{\Omega_i}\nabla Q(t)\text{d}V=\int_{\partial\Omega_i}Q(t)\boldsymbol{\tau}\text{d}\Gamma \\
&=\sum_{p\in N(i)} \Big( \sum_{m=1}^2 \omega_{p,m} \Big(\int\boldsymbol\psi f(\boldsymbol{x}_{p,m},t,\boldsymbol{u},\xi)\text{d}\Xi\Big) \boldsymbol{\tau} |\Gamma_{p}| \Big),
\end{split}
\end{equation*}
where $\Omega_i$ is arbitrary cell in computational mesh, $(\nabla Q)_i$ is the cell averaged gradient of the flow variable over 
cell $\Omega_i$, $\nabla Q$ is the distribution of flow gradient, $\boldsymbol{\tau}$ takes $\boldsymbol{n_x}$ and $\boldsymbol{n_y}$, which are the unit directions of local coordinate on the cell interface. 
The macroscopic variables $Q_{p,m}^{n+1}$ at a cell interface can be obtained by taking moments of $f_{p,m}^{n+1}$ and the cell interface values can be used for the reconstruction at the beginning of next time step. 
In order to utilize the two-stage fourth-order temporal discretization for the gas distribution function, the second-order gas-kinetic solver is needed.  
In the previous study, the gas distribution function is recommended to be updated by 
\begin{equation}\label{two-stage-3}
\begin{split}
&f^*=f^n+\frac{1}{2}\Delta tf_t^n, \\
&f^{n+1}=f^n+\Delta t f_{t}^*.
\end{split}
\end{equation}
where the subscripts $m,p$ are omitted. 
It is relatively simple to be implemented, but the numerical tests show that only third-order accuracy can be achieved.  
With Gaussian quadrature rule, the cell-averaged gradient $(\nabla Q)_i$ at $t=t^n$ can be obtained. 
It means that the computation of cell-averaged gradient $(\nabla Q)_i$ can be turned to calculate the point value $Q_{p,m}$ on the cell interface in the local orthogonal coordinate.

In the traditional HWENO method, the quadratic polynomial and linear polynomials need to be constructed at every time step. 
For the stationary mesh, the matrix for geometric information is stored at the initial step. 
However, for the ALE computation, with the moving mesh, 
the matrix needs to be calculated at every time step, which is very time consuming. 
To improve the efficiency, the memory-reduction compact reconstruction \cite{GKS-MR-compact} is introduced for unstructured meshes. 
For the target cell $\Omega_{i}$, the big stencil for the compact reconstruction is selected as 
\begin{align*}
  S_i&=\{Q_{i},Q_{i_1},\dots, Q_{i_p},\nabla Q_{i},\nabla Q_{i_1},\dots,\nabla Q_{i_p}\},
\end{align*}
where $Q_{i_p}$ and $\nabla Q_{i_p}$ are the cell averaged conservative variable and cell averaged gradient of the neighboring cell $\Omega_{i_p}$.
For simplicity, the big stencil is rearranged as 
\begin{align*}
  S_i&=\{Q_0,Q_1,…,Q_{p},\nabla Q_0,...,\nabla Q_{p}\}. 
\end{align*}
Based on the big stencil, a quadratic polynomial can be constructed as
\begin{align*}
  P^2(\boldsymbol{x})&=Q_{0}+\sum_{|\boldsymbol n|=1}^2a_{\boldsymbol n}p_{\boldsymbol n}(\boldsymbol{x}), 
\end{align*}
where $Q_{0}$ is the cell averaged conservative variables over $\Omega_{0}$, 
the multi-index $\boldsymbol n=(n_1, n_2, n_3)$ and $|\boldsymbol n|=n_1+n_2+n_3$. 
The base function $p_{\boldsymbol n}(\boldsymbol{x})$ is defined as
\begin{align*}
\displaystyle p_{\boldsymbol n}(\boldsymbol{x}) &= (x-x_0)^{n_1} (y-y_0)^{n_2}(z-z_0)^{n_3}\\
&-\frac{1}{\left|\Omega_{0}\right|}\iiint_{\Omega_{0}}(x-x_0)^{n_1}(y-y_0)^{n_2}(z-z_0)^{n_3}\text{d}V,
\end{align*}
where $(x_0,y_0,z_0)$ is the mass center of $\Omega_0$. 
To determine the coefficients of $P^2(\boldsymbol{x})$,  the derivatives are given by
\begin{align*}
\partial_x P^2(\boldsymbol{x})&=a_1 +a_{11}(x-x_0)+a_{12}(y-y_0)+a_{13}(z-z_0),\\
\partial_y P^2(\boldsymbol{x})&=a_2 +a_{12}(x-x_0)+a_{22}(y-y_0)+a_{23}(z-z_0),\\
\partial_z P^2(\boldsymbol{x})&=a_3 +a_{13}(x-x_0)+a_{32}(y-y_0)+a_{33}(z-z_0),
\end{align*}
where $\partial_x P^2(\boldsymbol{x}),\partial_y P^2(\boldsymbol{x}),\partial_z P^2(\boldsymbol{x})$ represent the derivatives in $x$, $y$, $z$ directions of $P^2(\boldsymbol{x})$. 
Meanwhile, the distribution of cell-averaged slopes can be written as linear polynomials as follows 
\begin{align*}
L^1_x(\boldsymbol{x})&=b_0 +b_1(x-x_0)+b_2(y-y_0)+b_3(z-z_0),\\
L^1_y(\boldsymbol{x})&=c_0 +c_1(x-x_0)+c_2(y-y_0)+c_3(z-z_0),\\
L^1_z(\boldsymbol{x})&=c_0 +d_1(x-x_0)+d_2(y-y_0)+d_3(z-z_0),
\end{align*}
where the following constraints are satisfied
\begin{equation*}
\begin{split}
\frac{1}{\left|\Omega_{k}\right|}\iiint_{\Omega_{k}}L^1_x(\boldsymbol{x})\text{d}V&=(Q_x)_{k},~(Q_{x})_{k}\in S_i, \\
\frac{1}{\left|\Omega_{k}\right|}\iiint_{\Omega_{k}}L^1_y(\boldsymbol{x})\text{d}V&=(Q_y)_{k},~(Q_{y})_{k}\in S_i,\\
\frac{1}{\left|\Omega_{k}\right|}\iiint_{\Omega_{k}}L^1_z(\boldsymbol{x})\text{d}V&=(Q_z)_{k},~(Q_{z})_{k}\in S_i.
\end{split}
\end{equation*}
With the least-square process, the coefficients in $L^1_x(\boldsymbol{x})$, $L^1_y(\boldsymbol{x})$,  $L^1_z(\boldsymbol{x})$ can be determined.
Compared with $\partial_x P^2(\boldsymbol{x})$, $\partial_y P^2(\boldsymbol{x})$, $\partial_z P^2(\boldsymbol{x})$, the quadratic terms in $P^2(\boldsymbol{x})$ can be obtained as 
\begin{align*}
a_{11} =  b_1,  a_{22} =  c_2, a_{33} = d_3,  
\end{align*}
and
\begin{align*}
a_{12} = \frac{b_2 +c_1}{2}, a_{13} = \frac{d_1 +b_3}{2}, a_{23} = \frac{d_2 +c_3}{2}.
\end{align*}
To fully determine the quadratic polynomial $P^2(\boldsymbol{x})$, the constraints for the cell averaged variable are considered
\begin{align*} 
\frac{1}{\left|\Omega_{k}\right|}\iint_{\Omega_{k}}P^2(\boldsymbol{x})\text{d}V&=Q_{k}, ~Q_{k}\in S_i. 
\end{align*}
Moving the quadratic terms to the right-hand side, a new linear system of the linear terms can be obtained
\begin{align*} 
\iiint_{\Omega_{k}}(\sum_{|\boldsymbol n|=1}a_{\boldsymbol n}p_{\boldsymbol n}(\boldsymbol{x}))\text{d}V& = 
\left|\Omega_{k}\right|\Delta Q_{k} - \iiint_{\Omega_{k}}(\sum_{|\boldsymbol n|=2}a_{\boldsymbol n}p_{\boldsymbol n}(\boldsymbol{x}))\text{d}V, 
\end{align*}
where $\Delta Q_{k}= (Q_{k}-Q_{0})$.  Solving the linear system with the least square method, all coefficients of $P^2(\boldsymbol{x})$ for the big stencil are determined.

\subsection{Limiting procedure}
The quadratic polynomial $P^2$ provides a desired third-order spatial accuracy in smooth regions. 
However, the unconstrained polynomial will generate numerical oscillations near discontinuities. 
To deal with discontinuities,  the gradient compression factor \cite{GKS-compact-gcf,GKS-MR-compact} 
and the generalized essentially non-oscillatory  (GENO) nonlinear reconstruction \cite{GKS-GENO} are adopted. 

In the limiting procedure,  a robust low-order polynomial needs to be constructed. 
For the target cell $\Omega_0$, a linear polynomial can be constructed based on the cell-averaged variables $\{Q_0,Q_1,…,Q_{p}\}$ as
\begin{equation*}
 P^1(\boldsymbol{x}) =Q_{0}+\sum_{|\boldsymbol n|=1}b_{\boldsymbol n}p_{\boldsymbol n}(\boldsymbol{x}), 
\end{equation*}
where the linear coefficients $b_n$ can be solved with one least-square process. 
To limit the linear polynomial $P^1(\boldsymbol{x})$ and avoid numerical oscillations, 
the gradient compression factor \cite{GKS-compact-gcf,GKS-MR-compact} is adopted. 
The compression factor $\phi$ in the target cell $\Omega_0$ is given as 
\begin{equation*}
\phi=\prod_{p\in N(i)} \prod_{m=1}^{2} \phi_{p,m}.
\end{equation*}
where $\phi_{p,m}$ denotes the compression factor for the $m$th Gaussian point at the $p$th interface, 
which can be calculated by 
\begin{align*}
\phi_{p,m}=\frac{1}{1+A^2}, 
\end{align*}
where
\begin{align*}
A=\frac{\left|p^l-p^r\right|}{p^l}+\frac{\left|p^l-p^r\right|}{p^r}+\left(\mathrm{Ma}_n^l-\mathrm{Ma}_n^r\right)^2+\left(\mathrm{Ma}_t^l-\mathrm{Ma}_t^r\right)^2, 
\end{align*}
where $p$ is pressure, $Ma_n$ and $Ma_t$ are the Mach numbers defined by normal and tangential velocity, 
and the superscripts $l, r$ denote the left and right values of the Gaussian points. 
Then the reconstructed linear polynomial $P^1(\boldsymbol{x})$ can be modified as
\begin{equation*}
  \widetilde P^1(\boldsymbol{x}) =Q_{0} + \phi \sum_{|\boldsymbol n|=1}b_{\boldsymbol n}p_{\boldsymbol n}(\boldsymbol{x}). 
\end{equation*}

To further improve  limiting procedure,  the GENO method \cite{GKS-GENO} is used for the nonlinear combination 
of the high-order polynomial $P^2(\boldsymbol{x})$ and robust linear polynomial $\widetilde P^1(\boldsymbol{x})$ as follows
\begin{equation*}
 R(\boldsymbol{x}) =  \chi P^2(\boldsymbol{x}) + (1-\chi) \widetilde P^1(\boldsymbol{x}),
\end{equation*}
where $R(\boldsymbol{x})$ is the final reconstructed polynomial. 
The GENO method adaptively transits from high-order polynomial in smooth regions to robust low-order polynomial near discontinuities with a path function $\chi$ as
The path function $\chi(\alpha)$ is defined as
\begin{equation*}
  \chi(\alpha)=\tanh(C\alpha)/\tanh(C),
\end{equation*}
where $C=20$ is a constant, $\alpha$ is the function of the smoothness indicator as 
\begin{align*}
\alpha= \frac{2\alpha^{H}}{\alpha^{H}+\alpha^{L}}, ~\alpha^{H} &= 1+(\frac{IS^{\tau}}{IS^{H}+\epsilon})^{r},~\alpha^{L}=1+(\frac{IS^{\tau}}{IS^{L}+\epsilon})^{r}. 
\end{align*}
where $\varepsilon = 10^{-16}$ is a small positive number, $IS^L$, $IS^H$ and $IS^\tau$ are the smoothness indicators
\begin{align*}
IS^{L}=\sum_{|\beta|=1}    h^{2|\beta|-3}
\int_{\Omega}  \left(D^{|\beta|}\widetilde P^1(\boldsymbol{x})\right)^{2}  \mathrm{d}V, \\
IS^{H}=\sum_{|\beta|=1}^2  h^{2|\beta|-3}
\int_{\Omega}  \left(D^{|\beta|}P^2(\boldsymbol{x})\right)^{2}  \mathrm{d}V, \\
IS^{\tau}=\sum_{|\beta|=2} h^{2|\beta|-3}
\int_{\Omega}  \left(D^{|\beta|}P^2(\boldsymbol{x})\right)^{2}  \mathrm{d}V,
\end{align*}
where $\beta=(\beta_1,\beta_2,\beta_3)$ is a multi-index with $|\beta|=\beta_1+\beta_2+\beta_3$.
Both GENO reconstruction and compression factor works well for the cases with moving meshes and strong shocks,
which verifies the robustness of current scheme.

\section{GPU implementation for high-order ALE method}

Recently, GPUs have been widely used for high-performance computing and provide great computational power for large-scale problems. 
In the previous studies, multi-GPU accelerated HGKS has been implemented on both structured and unstructured meshes for compressible flows \cite{GKS-GPU-str-2023,GKS-GPU-uns-2025}. 
In this paper, the high-order compact ALE scheme on unstructured meshes is implemented on multiple GPUs to accelerate the computation.

For single-GPU computation, 
the compute unified device architecture (CUDA) is used to implement the cell-level parallelism of the current ALE scheme. 
Computational procedures, including spatial reconstruction, flux evaluation, solution update and mesh movement, are implemented as separate CUDA kernels. 
The key to single-GPU acceleration is to define the kernels and assign the corresponding thread grids. 
With appropriate block and grid sizes, 
a one-to-one mapping between threads and mesh nodes, faces or cells can be established, enabling the fine-grained parallelism.

\begin{figure}[!h]
\centering
\includegraphics[width=0.95\textwidth]{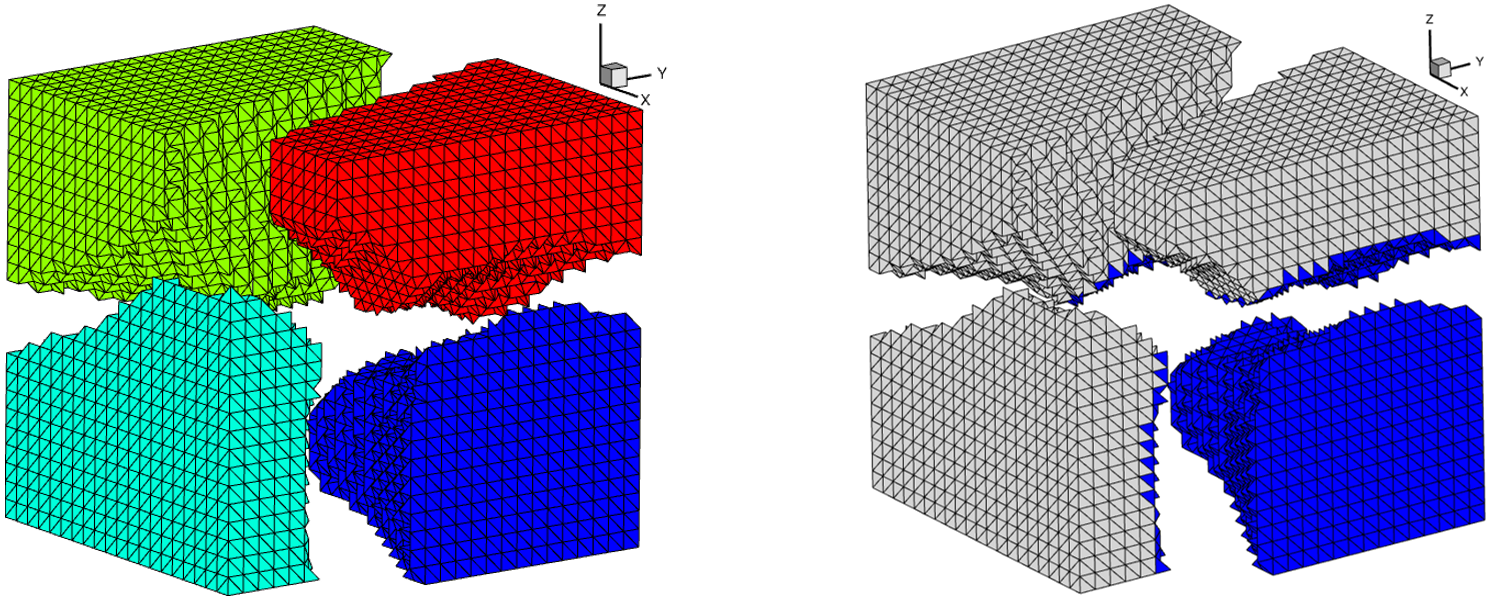}
\caption{\label{METIS-split} The unstructured Eulerian cells are split by METIS. }
\end{figure}

For multi-GPU computation, the unstructured mesh is first partitioned into subdomains using METIS \cite{METIS-Kar}, 
with each subdomain assigned to one GPU. 
As shown in Fig.\ref{METIS-split}, 
the local connectivity, ghost cells and communication mappings are constructed for each subdomain. 
MPI is used to exchange the required ghost-cell data between neighboring subdomains. 
In addition to the flow variables, 
the ALE computation requires the synchronization of nodal coordinates and mesh velocities. 
After these data exchanges, 
the same kernels in the single-GPU implementation can be reused for each GPU.

\section{Numerical tests}
In this section, the numerical tests for smooth flows and flows with strong discontinuity will be presented to validate our numerical scheme.

For the GPU computations, the  NVIDIA Tesla V100 GPU are used with NVIDIA CUDA
and CUDA-Aware MPI of NVIDIA HPC SDK 21.7. As reference, the CPU
codes run on the station with 2 Intel(R) Xeon(R) Gold 5120 CPUs using
OpenMP directives. The detailed parameters of  GPU are given
in Tab.\ref{GPU-CPU-B}, respectively.
There are $8$ GPUs inside one GPU node,
and more nodes are needed for more than $8$ GPUs. The GPU-GPU
communication in one GPU node is achieved by NVIDIA NVLink. The
communication across GPU nodes can be achieved by GPU Direct RDMA
via iWARP, RDMA over Converged Ethernet (RoCE) or InfiniBand. In
this paper, RoCE is used for communication across GPU nodes.

\begin{table}[!h]
\caption{\label{GPU-CPU-B} The detailed parameters of GPU.}
\begin{center}
\def\temptablewidth{0.6\textwidth}
{\rule{\temptablewidth}{1.0pt}}
        \begin{tabular*}{\temptablewidth}{@{\extracolsep{\fill}}c|c}
            ~                            &   NVIDIA Tesla V100 \\
            \hline
            Clock rate                      & 1.53 GHz          \\
            \hline
            Stream multiprocessor               & 80                \\
            \hline
            FP$32$ precision performance     & 15.7 Tflops       \\
            \hline
            FP$64$ precision performance    & 7834 Gflops       \\
            \hline
            GPU memory size                 & 32 GB             \\
            \hline
            Memory bandwidth                    & 897 GB/s          \\
        \end{tabular*}
        {\rule{\temptablewidth}{1.0pt}}
    \end{center}
\end{table}

The time step $\Delta t$ is given by
\begin{equation*}
\Delta t = \text{CFL} \cdot \mathop{\min_i}\left(\frac{\Delta r_i}{|\mathbf{U}_i|+a_i}\right),
\end{equation*}
where $i$ is the cell index, $\Delta r_i = \sqrt{|\Omega|_i}$ and $a_i$ is the sound speed in cell $i$. 
In the following numerical examples, the CFL number takes from $0.1$ to $0.3$. 
For the inviscid flow, the collision time $\tau$ takes 
\begin{align*}
\tau = \epsilon \Delta t+C\displaystyle|\frac{p^l-p^r}{p^l+p^r}|\Delta t,
\end{align*}
where $\varepsilon=0.1$ and $C=5$, $p^l$ and $p^r$ denote the pressure on the left and right sides of the cell interface. 
For the viscous flow, the collision time is related to the viscosity coefficient
\begin{align*}
\tau = \frac{\mu}{p} + C\displaystyle|\frac{p^l-p^r}{p^l+p^r}|\Delta t,
\end{align*}
where $\mu$ is the dynamic viscosity coefficient and $p$ is the pressure at the cell interface. 
Without spatial statement, the ratio of specific heats takes $\gamma=1.4$.

\begin{table}[!h]
\caption{\label{accuracy-static}
Advection of density perturbation: $L^1$ and $L^2$ errors and orders of accuracy with stationary meshes at $t=2$.}
\begin{center}
\def\temptablewidth{0.8\textwidth}
{\rule{\temptablewidth}{1.0pt}}
\begin{tabular*}{\temptablewidth}{@{\extracolsep{\fill}}c|cc|cc}
mesh            & $L^1$ error & Order & $L^2$ error & Order \\
\hline
Hex             &             &       &             &       \\
$10^3$          & 1.9108E-01  & ~     & 7.3799E-02  & ~     \\
$20^3$          & 2.7486E-02  & 2.7974 & 1.0723E-02  & 2.7829 \\
$40^3$          & 3.5098E-03  & 2.9692 & 1.3745E-03  & 2.9637 \\
$80^3$          & 4.4088E-04  & 2.9929 & 1.7267E-04  & 2.9927 \\
$160^3$         & 5.5168E-05  & 2.9985 & 2.1622E-05  & 2.9975 \\
\hline
Tet                     &             &       &             &       \\
$5^3  \times 6$ & 1.3923E-01  & ~     & 5.5436E-02  & ~     \\
$10^3 \times 6$ & 1.3498E-02  & 3.3666 & 5.3778E-03  & 3.3657 \\
$20^3 \times 6$ & 1.2554E-03  & 3.4265 & 5.0128E-04  & 3.4233 \\
$40^3 \times 6$ & 1.3418E-04  & 3.2259 & 5.3547E-05  & 3.2267 \\
$80^3 \times 6$ & 1.8123E-05  & 2.8883 & 7.6811E-06  & 2.8014 \\
\end{tabular*}
{\rule{\temptablewidth}{1.0pt}}
\end{center}
\end{table}

\begin{table}[!h]
\caption{\label{accuracy-type1}
Advection of density perturbation: $L^1$ and $L^2$ errors and orders of accuracy with Type-1 moving meshes at $t=2$.}
\begin{center}
\def\temptablewidth{0.7\textwidth}
{\rule{\temptablewidth}{1.0pt}}
\begin{tabular*}{\temptablewidth}{@{\extracolsep{\fill}}c|cc|cc}
mesh & $L^1$ error & order & $L^2$ error & order \\
\hline
Hex             &             &       &             &       \\
$10^3$          & 1.9651E-01  & ~      & 7.6923E-02  & ~      \\
$20^3$          & 3.0037E-02  & 2.7098 & 1.1820E-02  & 2.7022 \\
$40^3$          & 3.9480E-03  & 2.9275 & 1.5555E-03  & 2.9258 \\
$80^3$          & 4.9882E-04  & 2.9845 & 1.9651E-04  & 2.9847 \\
$160^3$         & 6.2499E-05  & 2.9966 & 2.4621E-05  & 2.9967 \\
\hline
Tet             &             &       &             &       \\
$5^3 \times 6$  & 1.4375E-01  & ~      & 5.7543E-02  & ~      \\
$10^3 \times 6$ & 1.5412E-02  & 3.2214 & 6.1963E-03  & 3.2151 \\
$20^3 \times 6$ & 1.4006E-03  & 3.4600 & 5.6877E-04  & 3.4455 \\
$40^3 \times 6$ & 1.4136E-04  & 3.3086 & 5.7505E-05  & 3.3061 \\
$80^3 \times 6$ & 1.8367E-05  & 2.9441 & 7.8154E-06  & 2.8793 \\
\end{tabular*}
{\rule{\temptablewidth}{1.0pt}}
\end{center}

\caption{\label{accuracy-type2} Advection of density perturbation: $L^1$ and $L^2$ errors and orders of accuracy with Type-2 moving meshes at $t=2$.}
\begin{center}
\def\temptablewidth{0.7\textwidth}
{\rule{\temptablewidth}{1.0pt}}
\begin{tabular*}{\temptablewidth}{@{\extracolsep{\fill}}c|cc|cc}
mesh            & $L^1$ error & Order & $L^2$ error & Order \\
\hline
Hex             &             &       &             &       \\
$10^3$          & 1.9406E-01  & ~      & 7.6124E-02  & ~      \\
$20^3$          & 2.8614E-02  & 2.7617 & 1.1323E-02  & 2.7491 \\
$40^3$          & 3.6973E-03  & 2.9522 & 1.4663E-03  & 2.9491 \\
$80^3$          & 4.6510E-04  & 2.9909 & 1.8555E-04  & 2.9823 \\
$160^3$         & 5.8267E-05  & 2.9968 & 2.3778E-05  & 2.9641 \\
\hline
Tet             &             &       &             &       \\
$5^3  \times 6$ & 1.4310E-01  & ~      & 5.7324E-02  & ~      \\
$10^3 \times 6$ & 1.3768E-02  & 3.3776 & 5.5767E-03  & 3.3616 \\
$20^3 \times 6$ & 1.2671E-03  & 3.4417 & 5.1318E-04  & 3.4419 \\
$40^3 \times 6$ & 1.3920E-04  & 3.1863 & 5.6133E-05  & 3.1926 \\
$80^3 \times 6$ & 1.6843E-05  & 3.0469 & 6.7915E-06  & 3.0471 \\
\end{tabular*}
{\rule{\temptablewidth}{1.0pt}}
\end{center}
\end{table}

% Efficiency1: MR3 .vs. WENO3

\begin{table}[!h]
\caption{\label{efficiency}
Efficiency comparison: the total execution time (s) and speedup for MR3-ALE and WENO3-ALE with advection of density perturbation at $t=2$.}
\begin{center}
\def\temptablewidth{0.85\textwidth}
\begin{tabular*}{\temptablewidth}{@{\extracolsep{\fill}}cccc}
\hline
Component                 & MR3     & WENO3    & Speedup \\
\hline
Total computation         & 2817.93 & 11097.60 & 3.94    \\
Least-squares computation & 708.63  & 2228.34  & 3.14    \\
Spatial reconstruction    & 205.26  & 7875.54  & 38.37   \\
Face evaluation and flux  & 1304.92 & 519.47   & 0.40    \\
Solution update           & 141.99  & 46.02    & 0.32    \\
\hline
\end{tabular*}
\end{center}
\end{table}

% Efficiency2: single-GPU .vs. CPU

\begin{table}[!h]
\caption{\label{gpu-cpu-efficiency}
Efficiency comparison: the detailed computational time (s) at $t=2$ and speedup for Tesla V100 GPU. }
\begin{center}
\def\temptablewidth{0.6\textwidth}
\begin{tabular*}{\temptablewidth}{@{\extracolsep{\fill}}cccc}
\hline
Mesh     & Tesla V100 & CPU     & Speedup \\
\hline
$10^3$   & 1.00       & 1.25    & 1.26    \\
$20^3$   & 2.75       & 14.36   & 5.23    \\
$40^3$   & 29.70      & 208.33  & 7.02    \\
$80^3$   & 449.75     & 2817.93 & 6.27    \\
\hline
\end{tabular*}
\end{center}
\end{table}

% Efficiency3: multi-GPU .vs. single-GPU
\begin{table}[!h]
\caption{\label{strong-scaling}
Efficiency test: the detailed computational time (s) at $t=2$ and parallel efficiency for strong scaling. }
\begin{center}
\begin{tabular}{ccc}
\hline
No. GPUs & $80^3$ & Parallel efficiency \\
\hline
1        & 449.75  & --   \\
2        & 215.10  & 1.05 \\
4        & 112.15  & 1.00 \\
8        & 60.54   & 0.93 \\
\hline
No. GPUs & $160^3$ & Parallel efficiency \\
\hline
1        & 6599.22 & --   \\
2        & 3359.66 & 0.98 \\
4        & 1703.78 & 0.97 \\
8        & 867.15  & 0.95 \\
\hline
\end{tabular}
\end{center}
\end{table}

% \begin{table}[!h]
% \caption{\label{weak-scaling}
% Efficiency test: the detailed computational time (s) at $t=2$ and parallel efficiency for weak scaling. }
% \begin{center}
% \begin{tabular}{cccc}
% \hline
% Mesh     & No. GPUs & Computational time & Ratio \\
% \hline
% $40^3$   & 1        & 29.70              & --    \\
% $80^3$   & 8        & 60.54              & 2.04  \\
% $80^3$   & 1        & 449.75             & --    \\
% $160^3$  & 8        & 867.15             & 1.93  \\
% \hline
% \end{tabular}
% \end{center}
% \end{table}

\begin{table}[!h]
\caption{\label{weak-scaling}
Efficiency test: the computational time (s) with $1600$ time steps and time ratio for weak scaling. }
\begin{center}
\begin{tabular}{cccc}
\hline
Mesh     & No. GPUs & Computational time & Ratio \\
\hline
$40^3$   & 1        & 118.80             & --    \\
$80^3$   & 8        & 121.08             & 1.02  \\
$80^3$   & 1        & 899.50             & --    \\
$160^3$  & 8        & 867.15             & 0.96  \\
\hline
\end{tabular}
\end{center}
\end{table}

\subsection{Accuracy tests}
In this case, the three-dimensional advection of density perturbation is used to test the order of accuracy with hexahedral and tetrahedral meshes. 
The computational domain is $[0,2]\times[0,2]\times[0,2]$, and the initial condition is set as 
\begin{align*}
\rho_0&(x, y, z)=1+0.2\sin(\pi(x+y+z)),~p_0(x,y,z)=1,\\
&U_0(x,y,z)=1,~V_0(x,y,z)=1,~W_0(x,y,z)=1.
\end{align*}
The periodic boundary conditions are adopted at all boundaries, and the analytic solution is given by
\begin{align*}
\rho(x,y&,z,t)=1+0.2\sin(\pi(x+y+z-3t)),~p(x,y,z,t)=1,\\
&U(x,y,z,t)=1,~V(x,y,z,t)=1,~W(x,y,z,t)=1.
\end{align*}
To test the accuracy for current reconstruction, the stationary uniform meshes with mesh size $h = 2/N$ are used. 
The $L^1$ and $L^2$ errors and orders at $t=2$ are presented in Tab.\ref{accuracy-static}, 
and the expected order of accuracy can be achieved on both hexahedral and tetrahedral meshes. 
To validate the order of accuracy with moving-meshes, two types of prescribed moving meshes are adopted.
At the initial time $t=0$, the hexahedral and tetrahedral meshes are uniformly distributed in $[0,2]\times[0,2]\times[0,2]$ with the mesh size $h=2/N$, 
and the initial coordinates for node $i$ is denoted by $(\xi_i,\eta_i,\zeta_i)$. 
The position of node $i$ at time $t$ is given by
\begin{align}
\text{Type 1}: &\begin{cases}
\displaystyle x_i = \xi_i   + 0.05 \sin(\pi \xi_i) \sin(\pi \eta_i) \sin(\pi \zeta_i) \sin(\pi t),\\
\displaystyle y_i = \eta_i  + 0.05 \sin(\pi \xi_i) \sin(\pi \eta_i) \sin(\pi \zeta_i) \sin(\pi t),\\
\displaystyle z_i = \zeta_i + 0.05 \sin(\pi \xi_i) \sin(\pi \eta_i) \sin(\pi \zeta_i) \sin(\pi t),
\end{cases}\label{mesh-type1}\\
\text{Type 2}: &\begin{cases}
\displaystyle x_i = \xi_i   + 0.05\sin (\pi \xi_i)   \sin(\pi t),\\
\displaystyle y_i = \eta_i  + 0.05\sin (\pi \eta_i)  \sin(\pi t),\\
\displaystyle z_i = \zeta_i + 0.05\sin (\pi \zeta_i) \sin(\pi t),
\end{cases}\label{mesh-type2}
\end{align}
The $L^1$ and $L^2$ errors and orders of accuracy at $t=2$ are shown in Tab.\ref{accuracy-type1} and Tab.\ref{accuracy-type2}, 
where the expected orders of accuracy are achieved.

\begin{figure}[!h]
\centering
\includegraphics[width=0.65\textwidth]{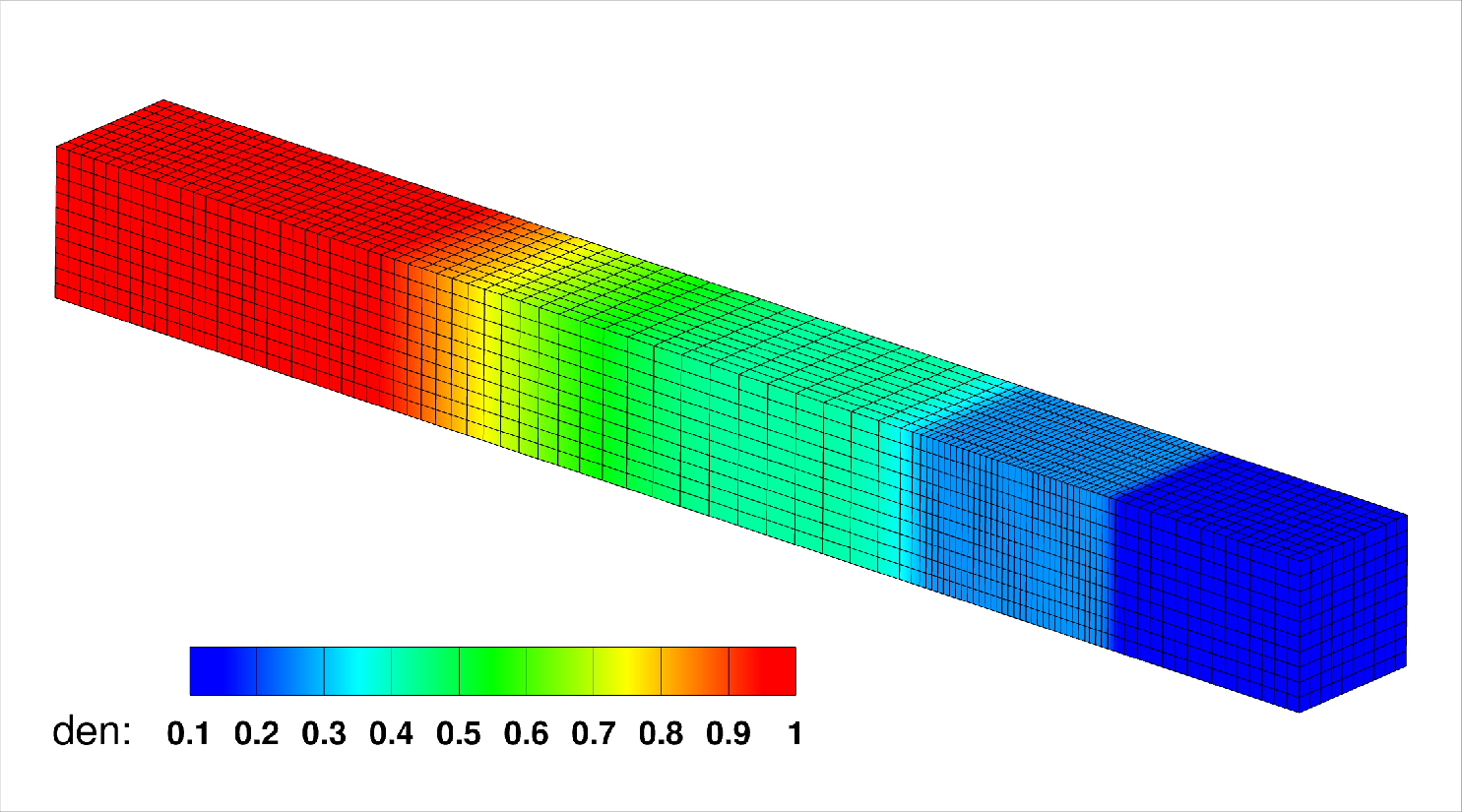}\\
\includegraphics[width=0.65\textwidth]{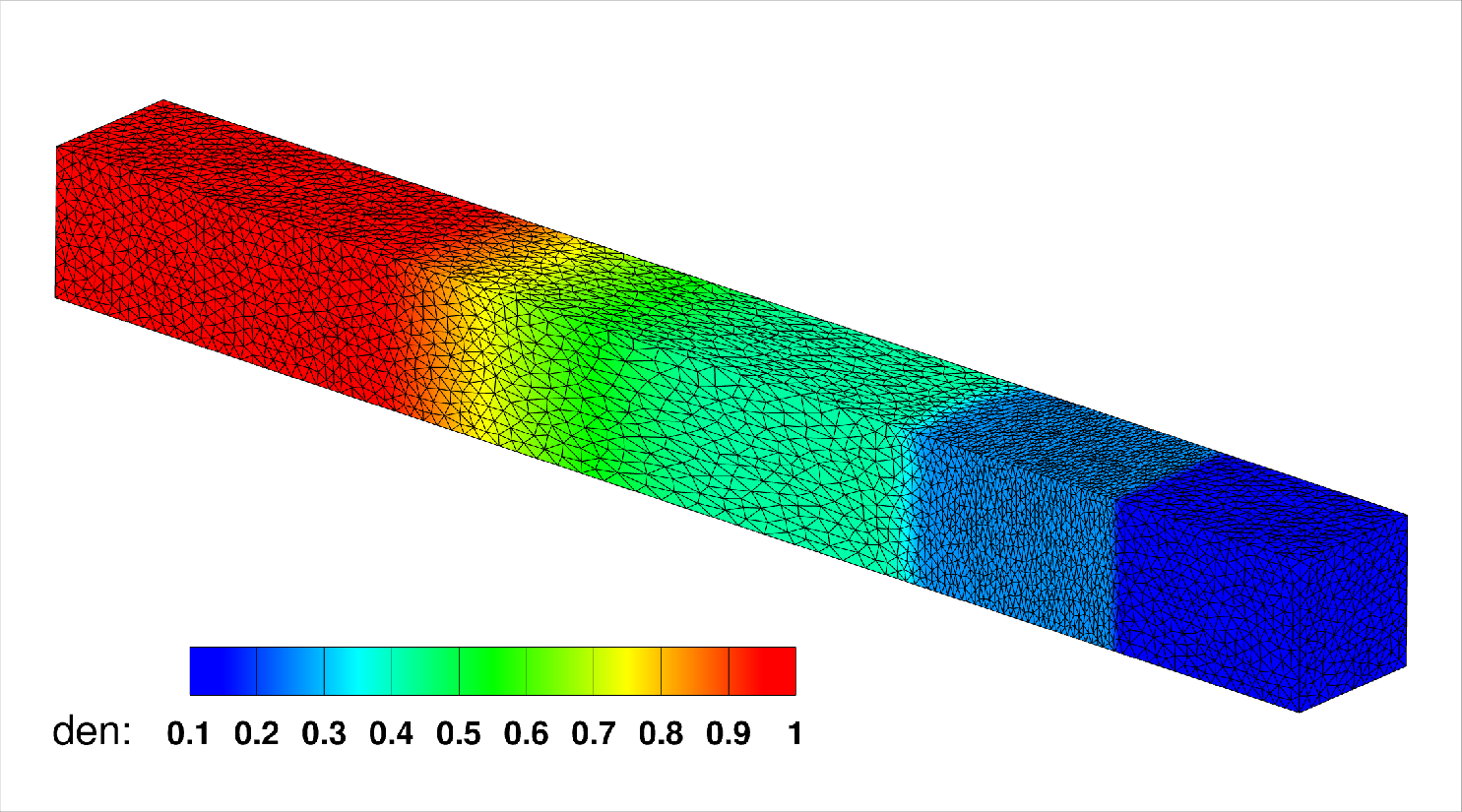}
\caption{\label{sod-flow} Sod problem: the density and mesh distributions with $h=1/100$ hexahedral (left) and tetrahedral (right) meshes at $t=0.2$.}
\end{figure}

\subsection{Efficiency test}
In this section,  the computational efficiency of the current scheme is evaluated. 
The first test compares performance of the current MR3-ALE scheme with the previous WENO3-ALE scheme \cite{GKS-ALE5}. 
The advection of density perturbation with the Type-1 moving mesh is adopted to compare the execution time. 
Both codes are run on an AMD Ryzen 7 8845HS processor with 8 cores and 16 threads at a base frequency of 3.8 GHz, using 16 OpenMP threads.
The execution time is measured by directly timing how long the code runs until $t = 2$. 
For the moving mesh problems, the geometric information needs to be computed at each time step, 
which needs a large number of matrix operations. 
In the previous WENO3-ALE scheme, the matrix operations are needed on both large and small stencils, 
which introduces significant computational cost. 
With the current memory-reduction reconstruction process, 
only one $3 \times 3$ coefficient matrix has to be regenerated inside each cell, and it is trivial to get the inverse matrix. 
The detailed computational times for MR3-ALE and WENO3-ALE on the $80^3$ mesh are presented in Tab.\ref{efficiency}.
Compared with WENO3-ALE scheme, MR3-ALE scheme achieves speedups of $3.14$ and $38.37$ for the least-squares computation and spatial reconstruction.
Although MR3-ALE costs more time in the face evaluation and solution updating process, 
the reduction in reconstruction cost results in an overall 4x speedup,  which validates the efficiency of the current scheme. 

To evaluate the performance of the GPU implementation, 
the advection of density perturbation with the Type-1 moving mesh is run with NVIDIA Tesla V100 GPUs. 
The CPU code using OpenMP is taken as a reference, 
and the speedup is defined as 
\begin{equation*}
  \text{Speedup} = \frac{ \text{CPU time} }{ \text{GPU time} }. 
\end{equation*}
The speedups are given in Tab.\ref{gpu-cpu-efficiency}, 
where the total execution times are measured at $t = 2$. 
As shown in Tab.\ref{gpu-cpu-efficiency}, a single Tesla V100 GPU achieves speedups of $5.23-7.02$ for $20^3$ to $80^3$ meshes. 
The multiple-GPU accelerated code is implemented as well, 
and the scalability is defined as 
\begin{equation*}
  \text{Parallel efficiency} = \frac{ \text{single-GPU time} }{ \text{multiple-GPU time} \times \text{Number of GPUs} }. 
\end{equation*}
The detailed parallel efficiency for strong scaling is given in Tab.\ref{strong-scaling}, 
where 1 to 8 GPUs are used for the case with $80^3$ and $160^3$ meshes. 
As can be seen, the multiple-GPU code scales properly with the increasing number of GPUs. 
To analyze the parallel efficiency for weak scaling, 
the numbers of cells and GPUs are both increased by a factor of eight, 
and the number of cells per GPU remains unchanged. 
The computational times are the total execution time with $1600$ time steps for all meshes.
As shown in Tab.\ref{weak-scaling},
the time ratios are close to unity, demonstrating good weak-scaling performance of current code.

\begin{figure}[!h]
\centering
\includegraphics[width=0.475\textwidth]{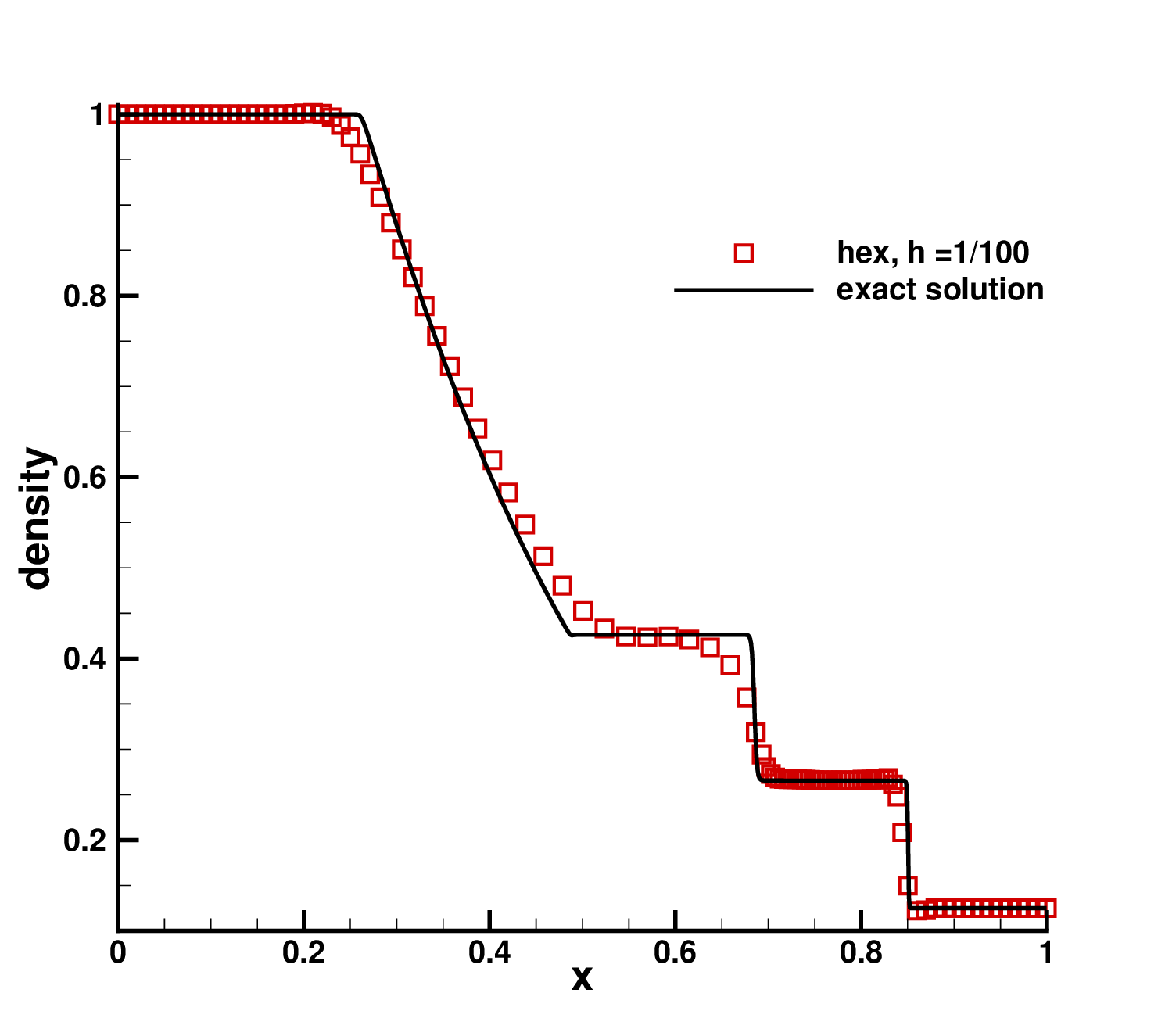}
\includegraphics[width=0.475\textwidth]{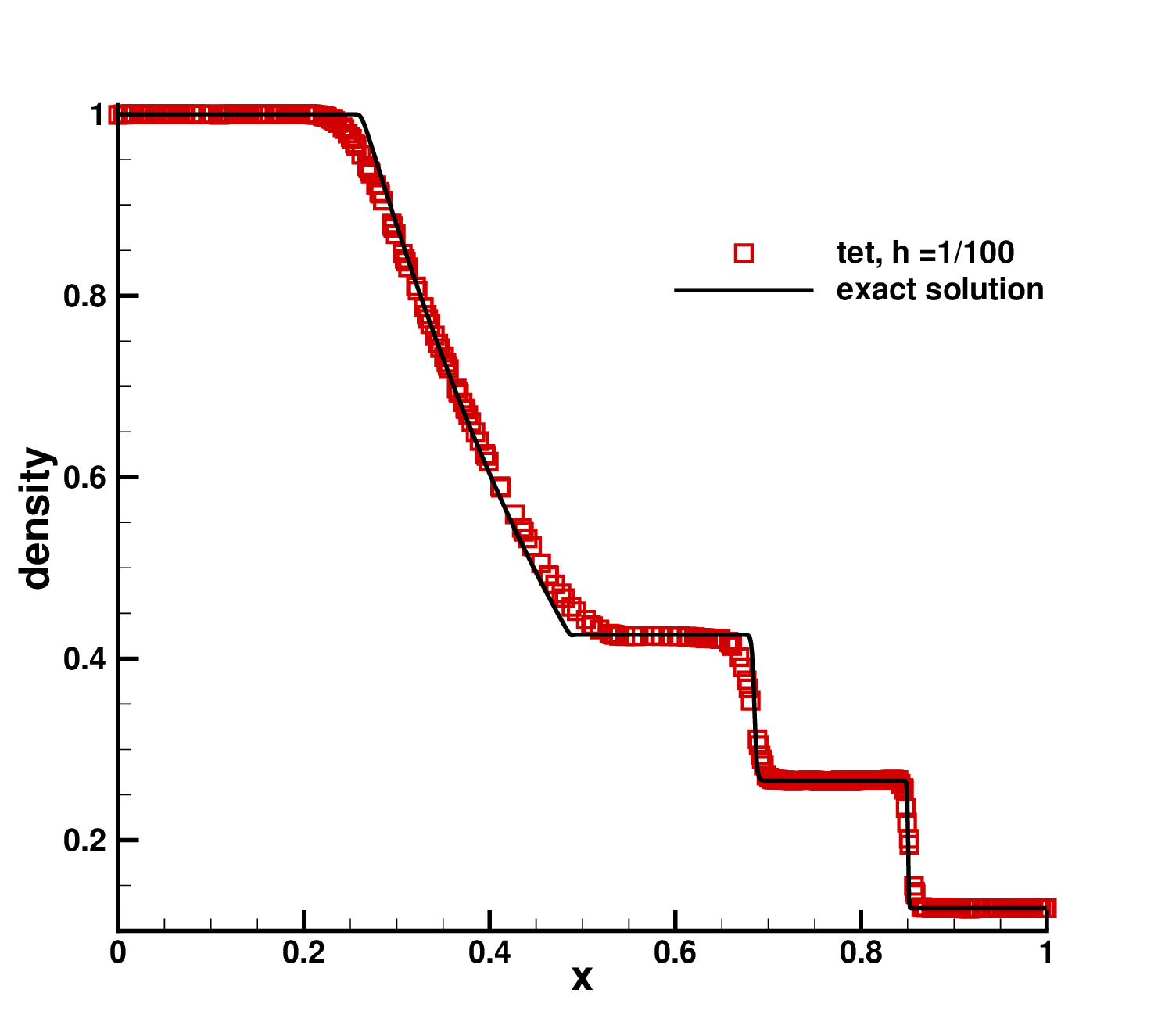}
\includegraphics[width=0.475\textwidth]{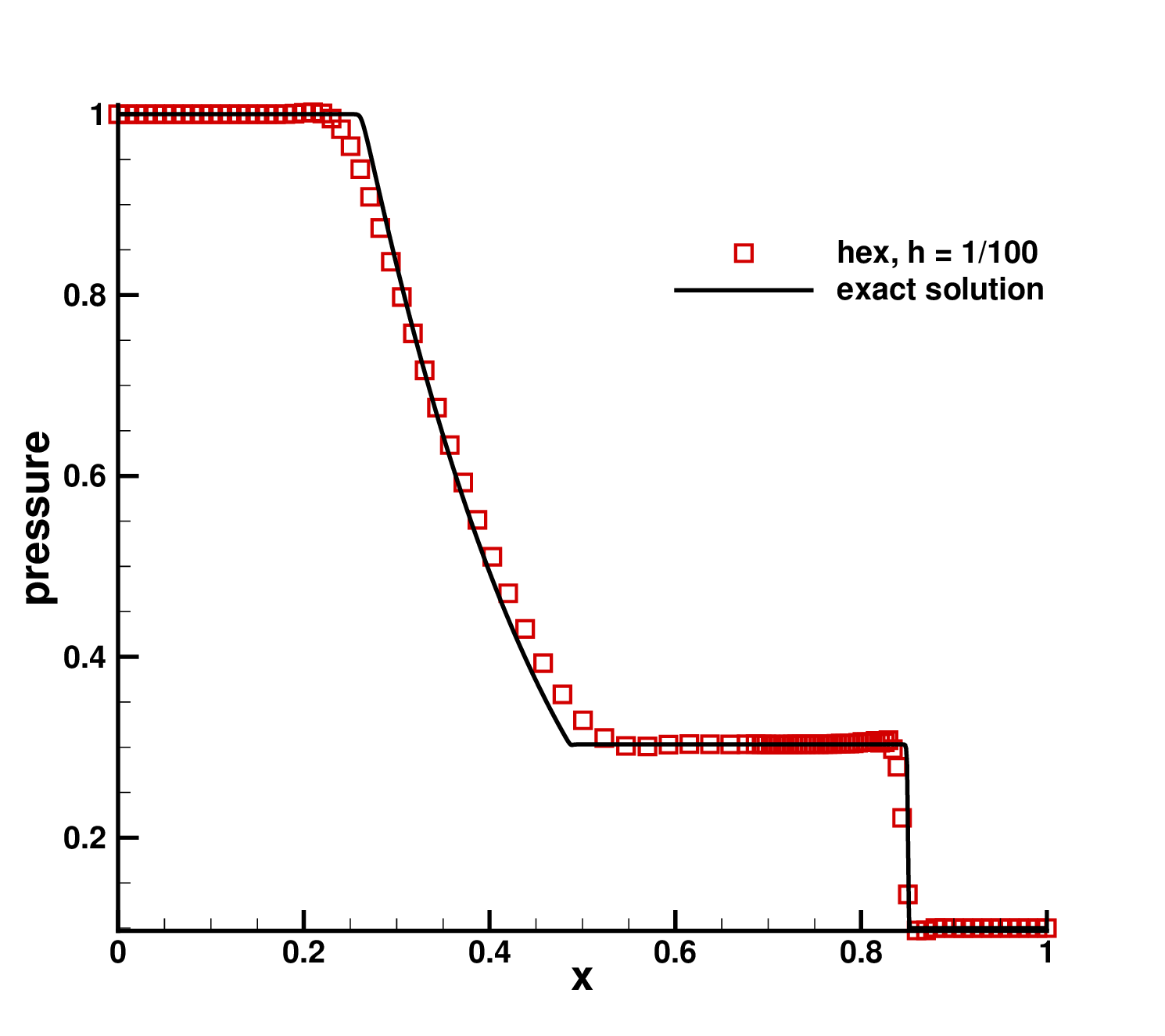}
\includegraphics[width=0.475\textwidth]{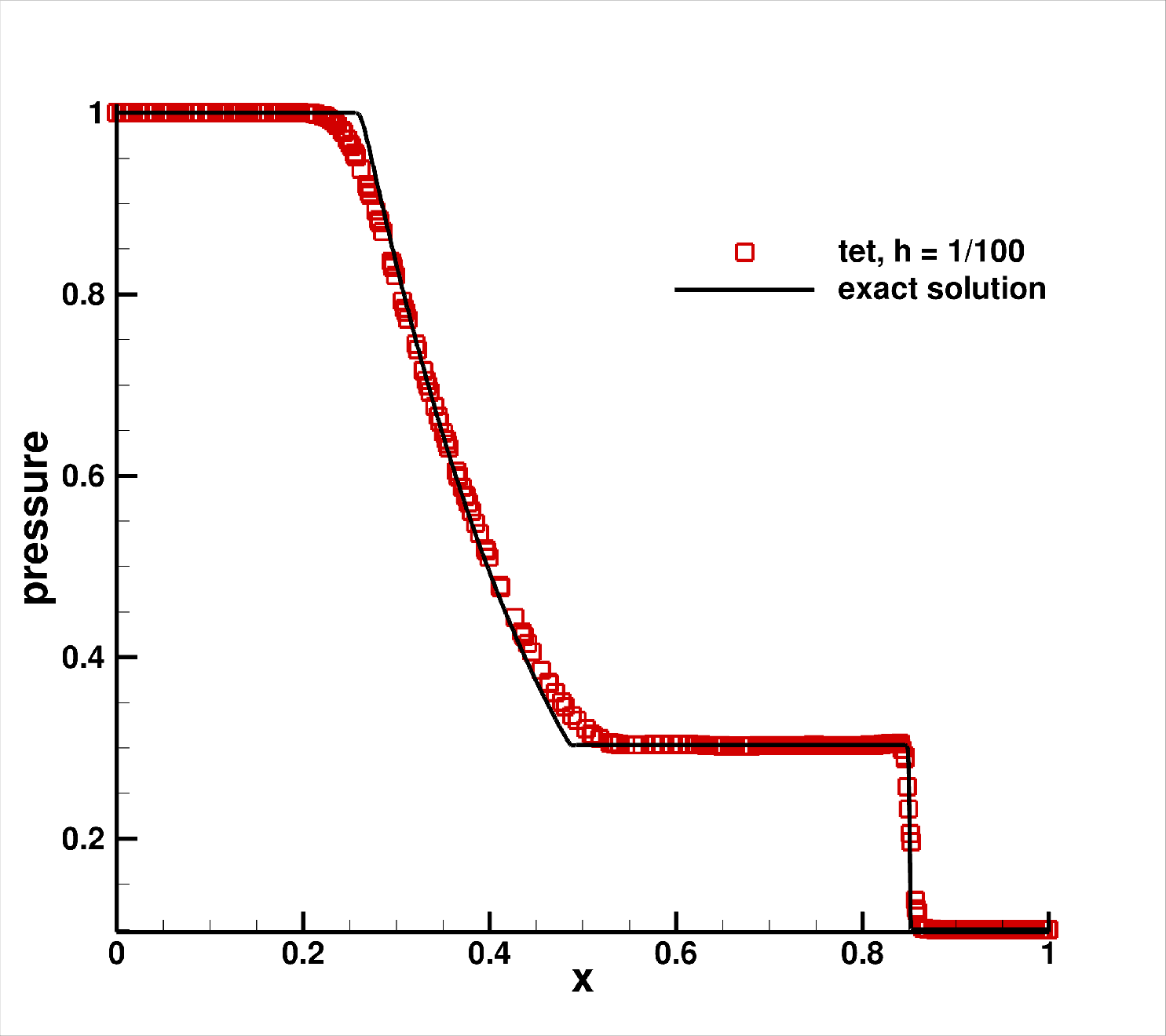}
\caption{\label{sod-data} Sod problem: the density (top) and pressure (bottom) profiles along the centerline 
with the hexahedral mesh (left) and tetrahedral mesh (right) at $t=0.2$. }
\end{figure}

\subsection{One-dimensional Riemann problem}
In this case, the Sod problem is tested, and the initial condition is given by
\begin{equation*}
(\rho, U, V, W, p)=
\begin{cases}
(1, 0, 0, 0, 1),  \ \ \ \ &  0\leq x \leq0.5,\\
(0.125, 0, 0, 0, 0.1),    &  0.5< x \leq1.
\end{cases}
\end{equation*}
The computational domain is $[0,1]\times[0,0.1]\times[0,0.1]$.
Initially, the uniform hexahedral mesh and the tetrahedral mesh with cell size $h=1/100$ are used in the computation.
The reflective boundary condition is applied at all boundaries. 
The mesh velocity is given by the cell-centered Lagrangian nodal solver \cite{Lagrangian-c1}, 
and the smooth procedure is used every $20$ iteration steps. 
The mesh and density distributions at $t=0.2$ are presented in Fig.\ref{sod-flow} for both hexahedral and tetrahedral meshes,
where the mesh movement can be clearly observed. 
The density and pressure distributions along the centerline at $t=0.2$ are shown in Fig.\ref{sod-data}. 
As can be seen, both the hexahedral and tetrahedral solutions agree well with the exact solutions. 
With the Lagrangian nodal velocity,  the mesh nodes automatically cluster near the shock, 
leading to a sharper resolution of the right-moving shock.

\begin{figure}[!h]
\centering
\includegraphics[width=0.475\textwidth]{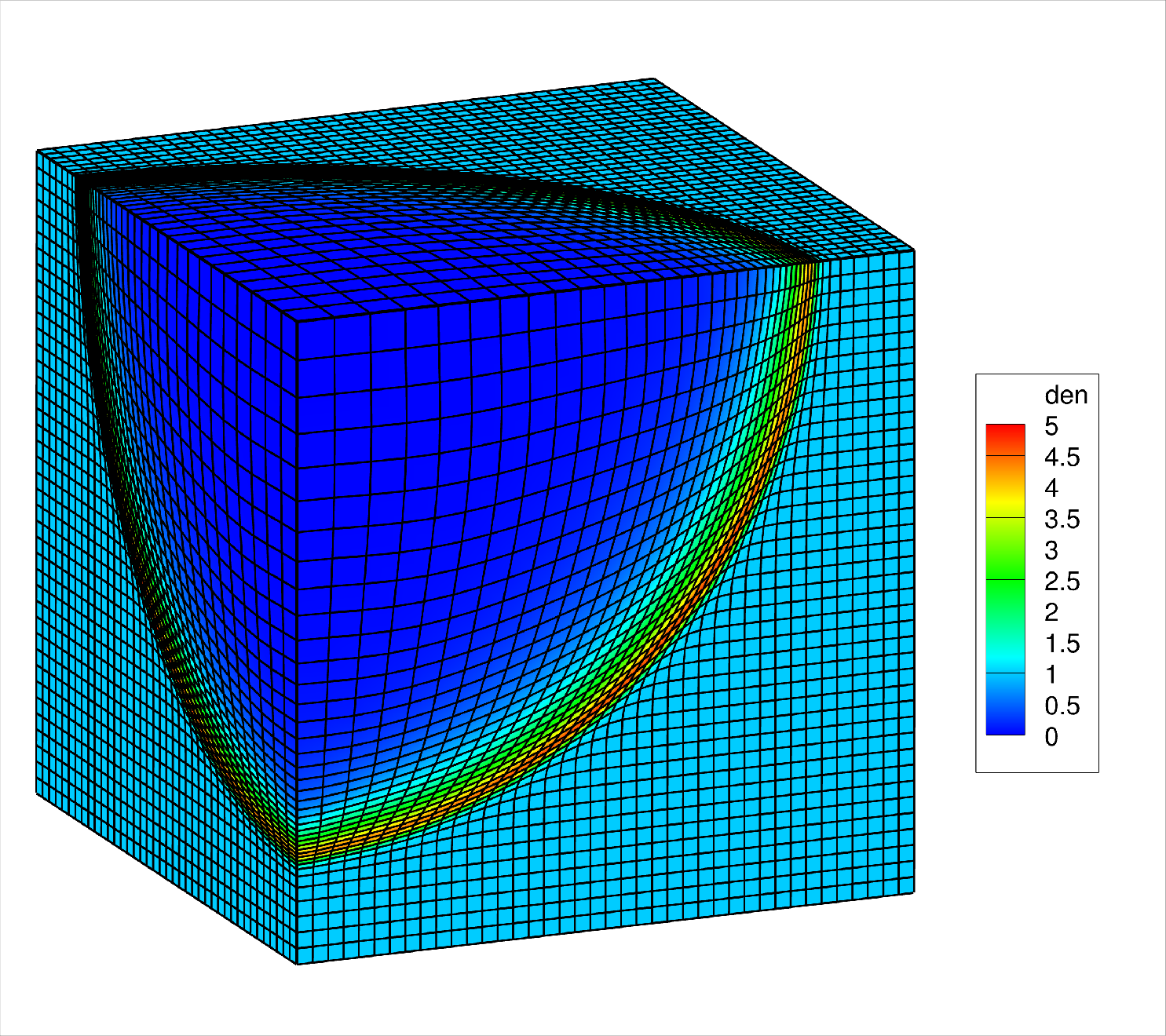}
\includegraphics[width=0.475\textwidth]{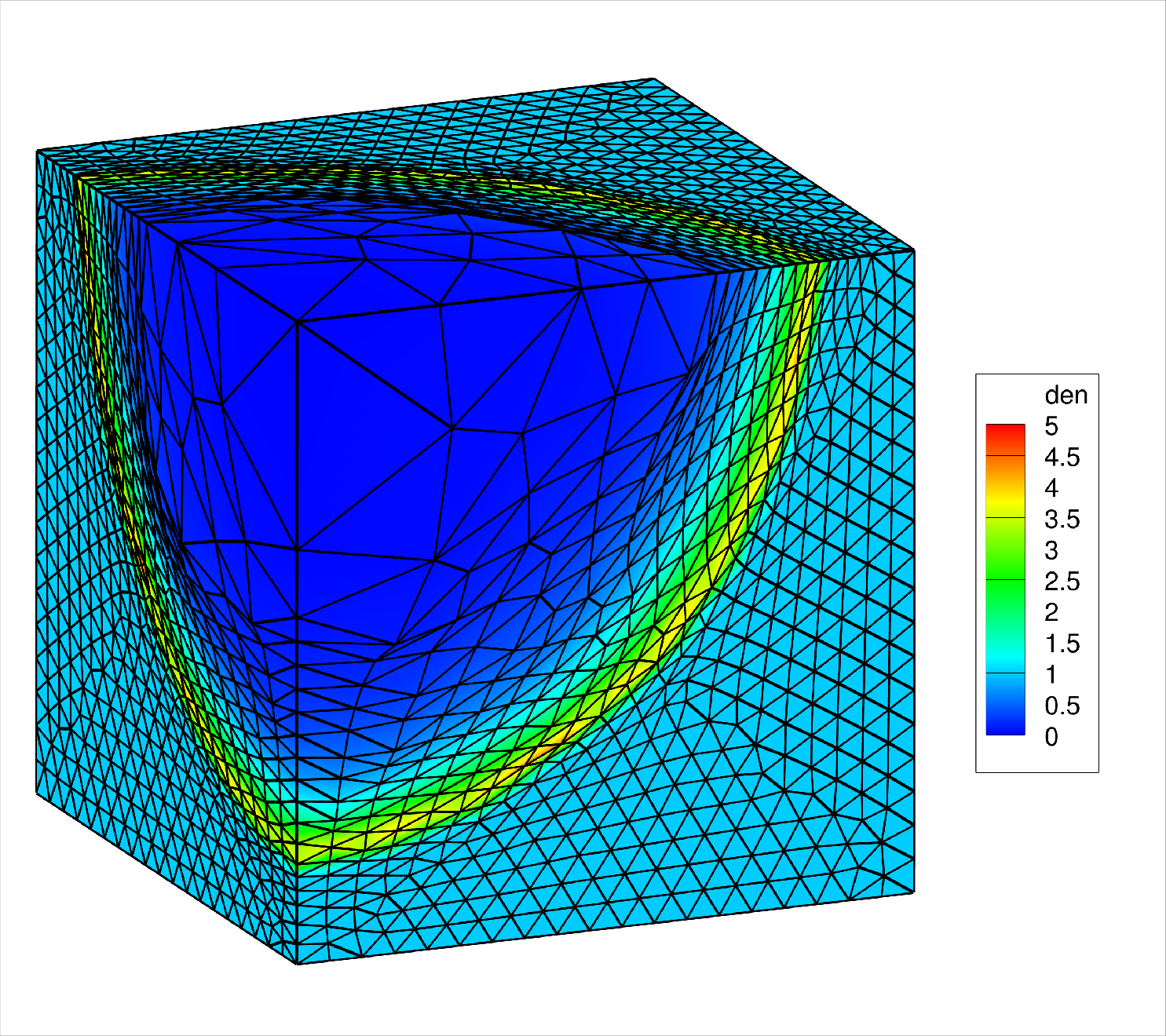}
\caption{\label{sedov-flow} Sedov problem: the density distribution with $h=1/30$ hexahedral mesh (left) and $h = 1/15$ tetrahedral mesh (right) at $t=1$.}
\end{figure}

\begin{figure}[!h]
\centering
\includegraphics[width=0.475\textwidth]{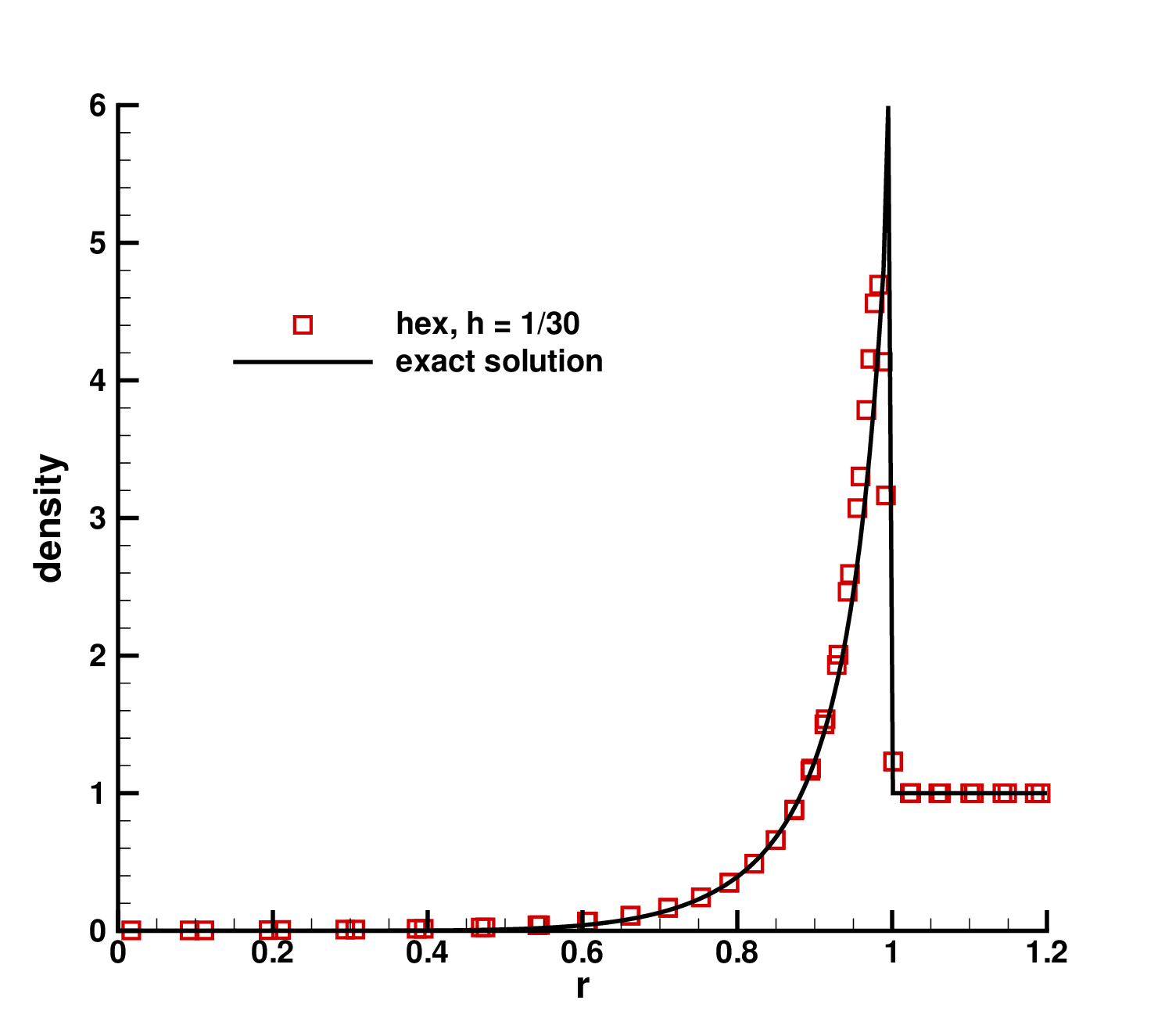} 
\includegraphics[width=0.475\textwidth]{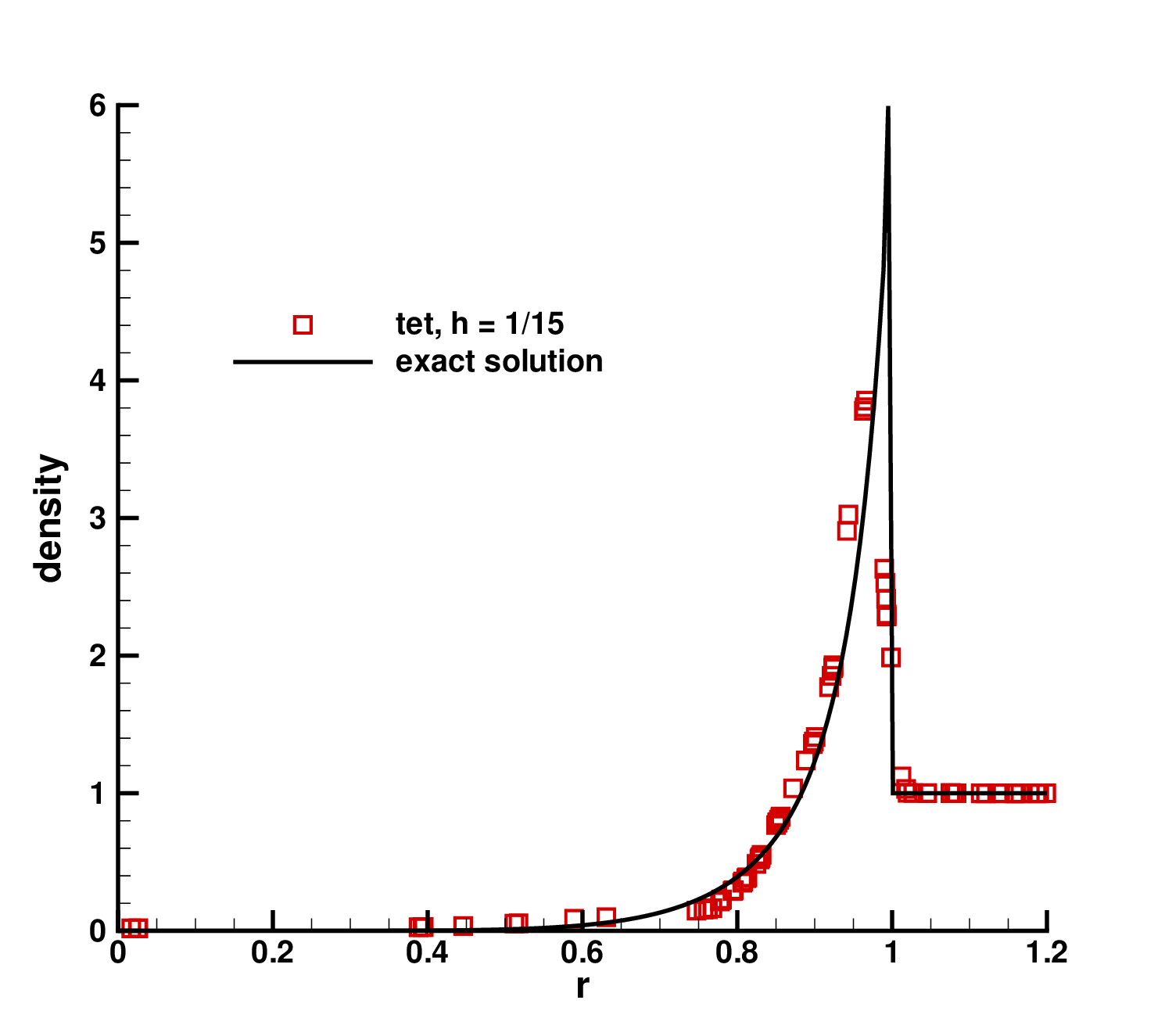} \\
\includegraphics[width=0.475\textwidth]{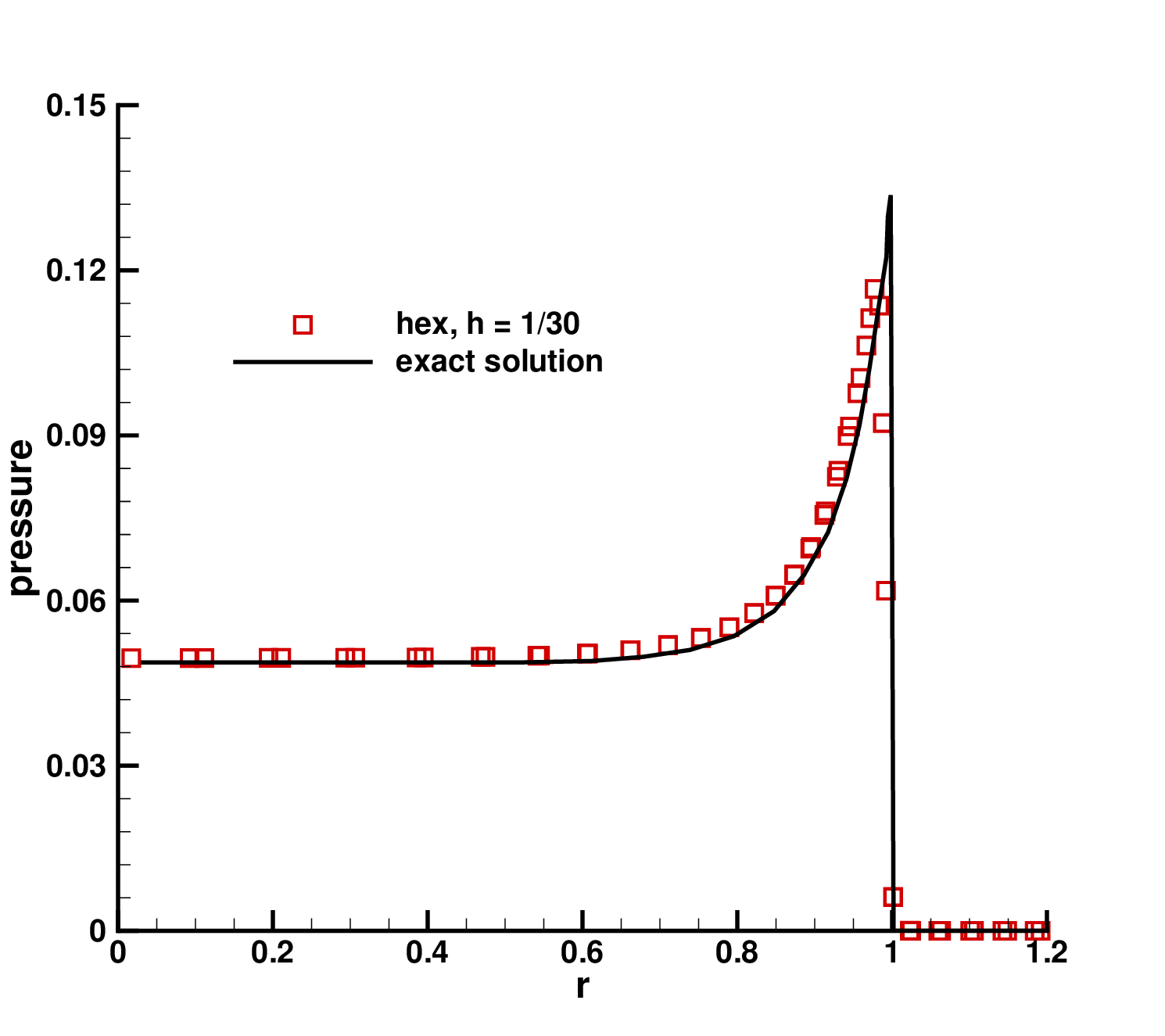} 
\includegraphics[width=0.475\textwidth]{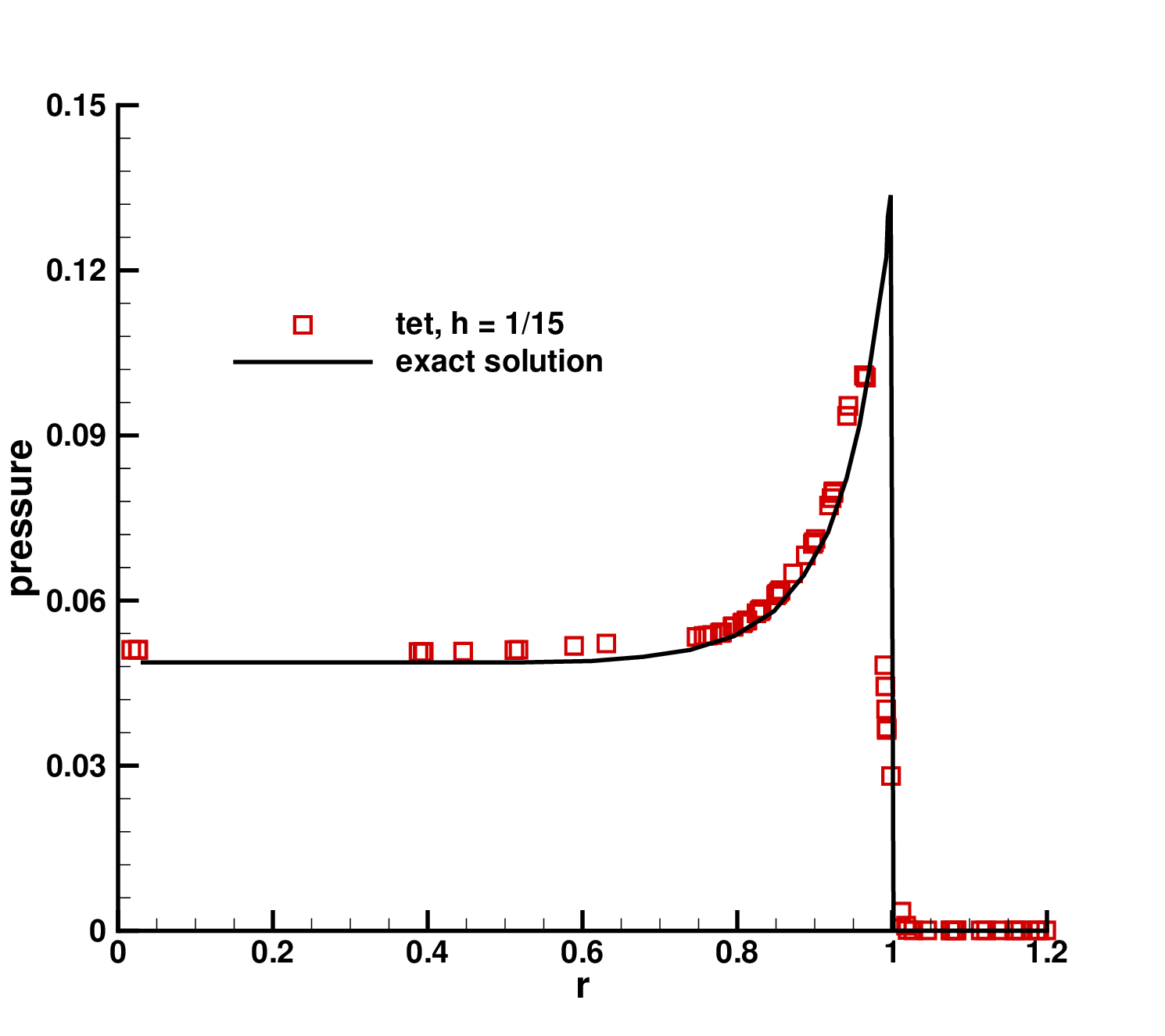} 
\caption{\label{sedov-data} Sedov problem: the density (top) and pressure (bottom) profiles along the diagonal line with 
$h=1/30$ hexahedral mesh (left) and $h=1/15$ tetrahedral mesh (left) at $t=1$.}
\end{figure}

\subsection{Sedov problem}

The Sedov problem describes a blast wave generated by energy deposition at a singular point \cite{Case-Sedov}, 
and is a benchmark problem for Lagrangian methods. 
The computational domain is $[0,1.2]^3$. 
The initial condition is given as $(\rho, U, V, W, p) = (1, 0, 0, 0, 10^{-6})$, except in the cells containing the origin. 
The pressure in the cells containing the origin is $p=(\gamma-1)\varepsilon_0/V$, 
where $\varepsilon_0=0.106384$ is the total released energy and $V$ is the total volume of those cells. 
The reflective boundary condition is imposed on the boundaries at $x=0$, $y=0$ and $z=0$ planes, 
and the non-reflective condition is used on the other boundaries. 
In the computation, the uniform $h=1/30$ hexahedral mesh and $h=1/15$ tetrahedral mesh are used initially. 
The mesh velocity is determined by the Lagrangian nodal solver \cite{Lagrangian-c1}, 
and the smoothing process is applied every $10$ iteration steps for both hexahedral and tetrahedral meshes. 
According to the self-similar solution \cite{Case-Sedov}, 
the solution at $t=1$ consists of a diverging infinite strength shock whose front is located at radius $r=1$. 
The density and mesh distributions at $t=1$ are given in Fig.\ref{sedov-flow}. 
It can be observed that the mesh follows the moving wavefront and is strongly compressed near the front, 
which poses a stringent requirement on the robustness of the numerical scheme. 
The density and pressure profiles along the diagonal line at $t=1$ are also given in Fig.\ref{sedov-data}. 
The numerical results agree well with the exact solution even with relatively coarse meshes, 
validating the robustness and accuracy of current scheme.

\begin{figure}[!h]
\centering
\includegraphics[width=0.6\textwidth]{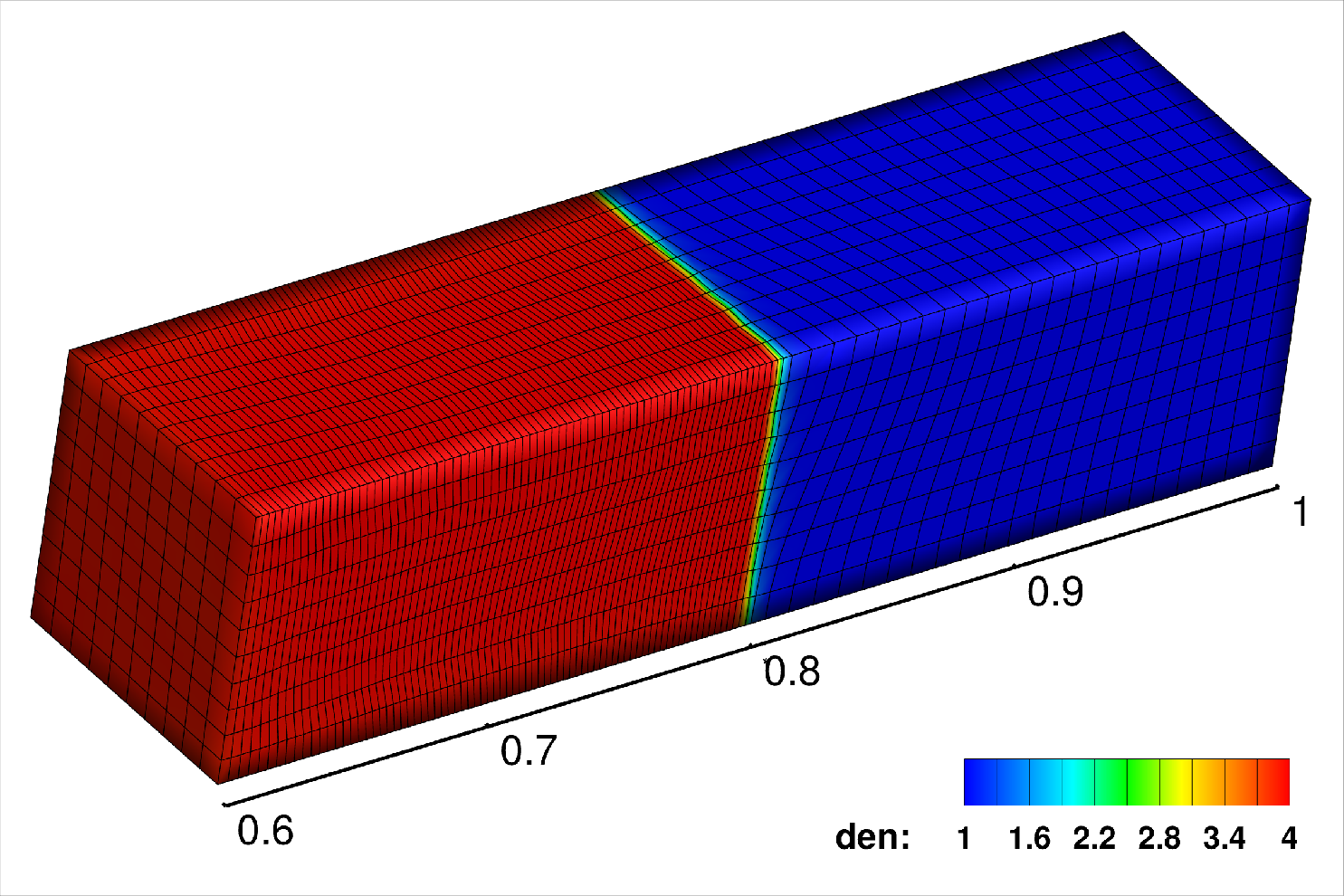}\\ 
\includegraphics[width=0.6\textwidth]{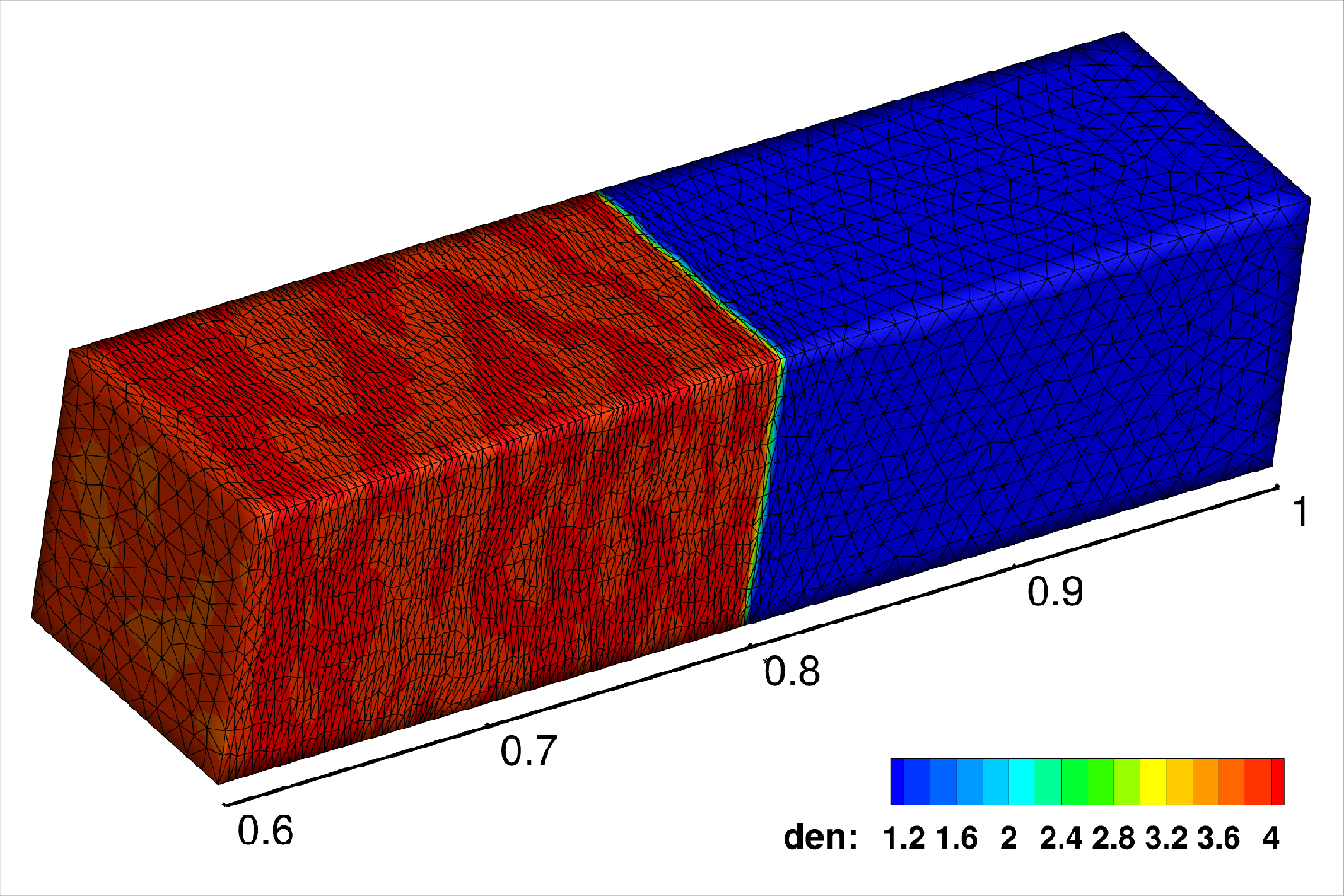}
\caption{\label{saltz-flow} Saltzman problem: the mesh and density distributions with $h=1/100$ hexahedral mesh (left) and tetrahedral mesh (right) at $t=0.6$. }
\end{figure}

\begin{figure}[!h]
\centering
\includegraphics[width=0.475\textwidth]{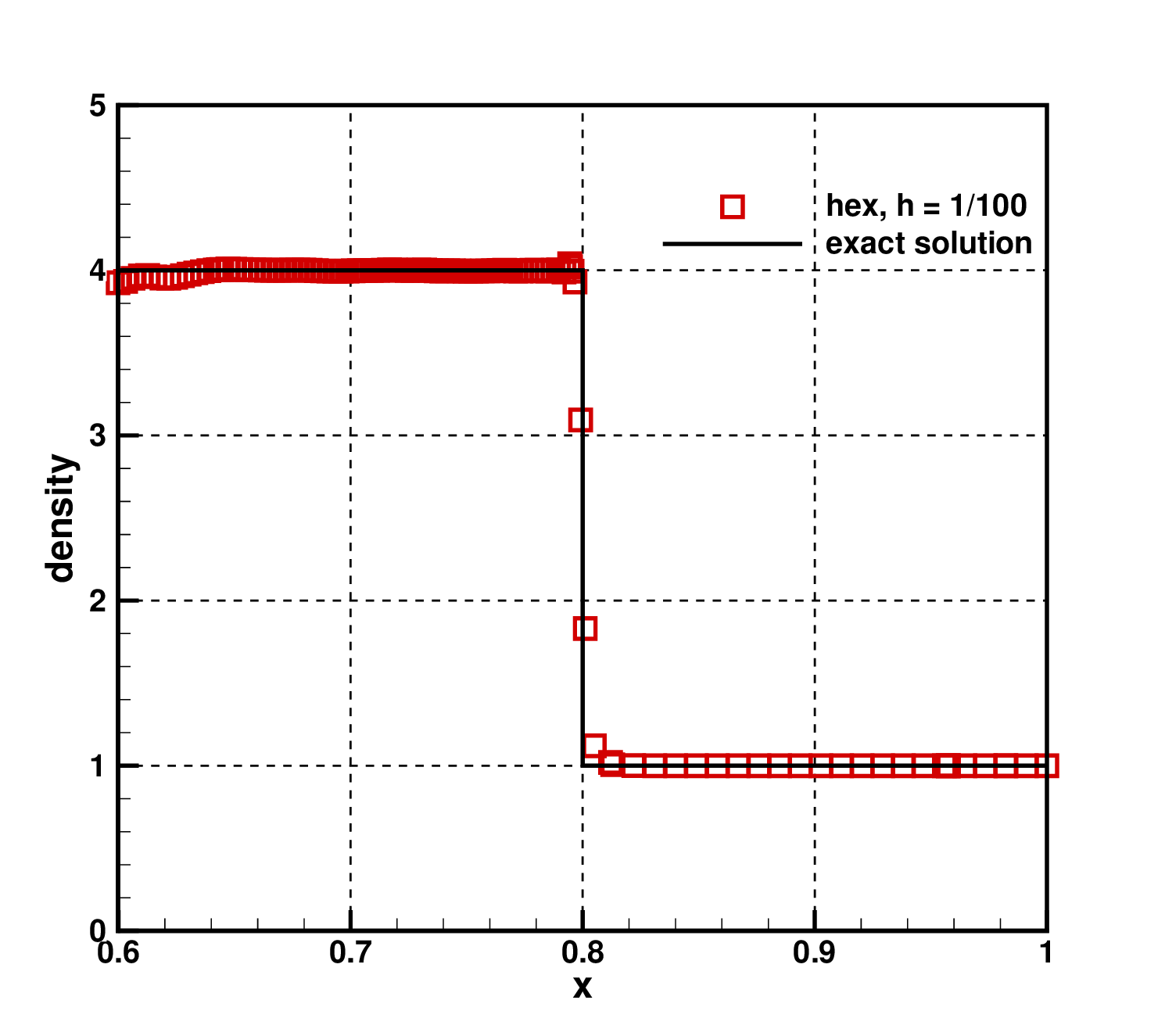} 
\includegraphics[width=0.475\textwidth]{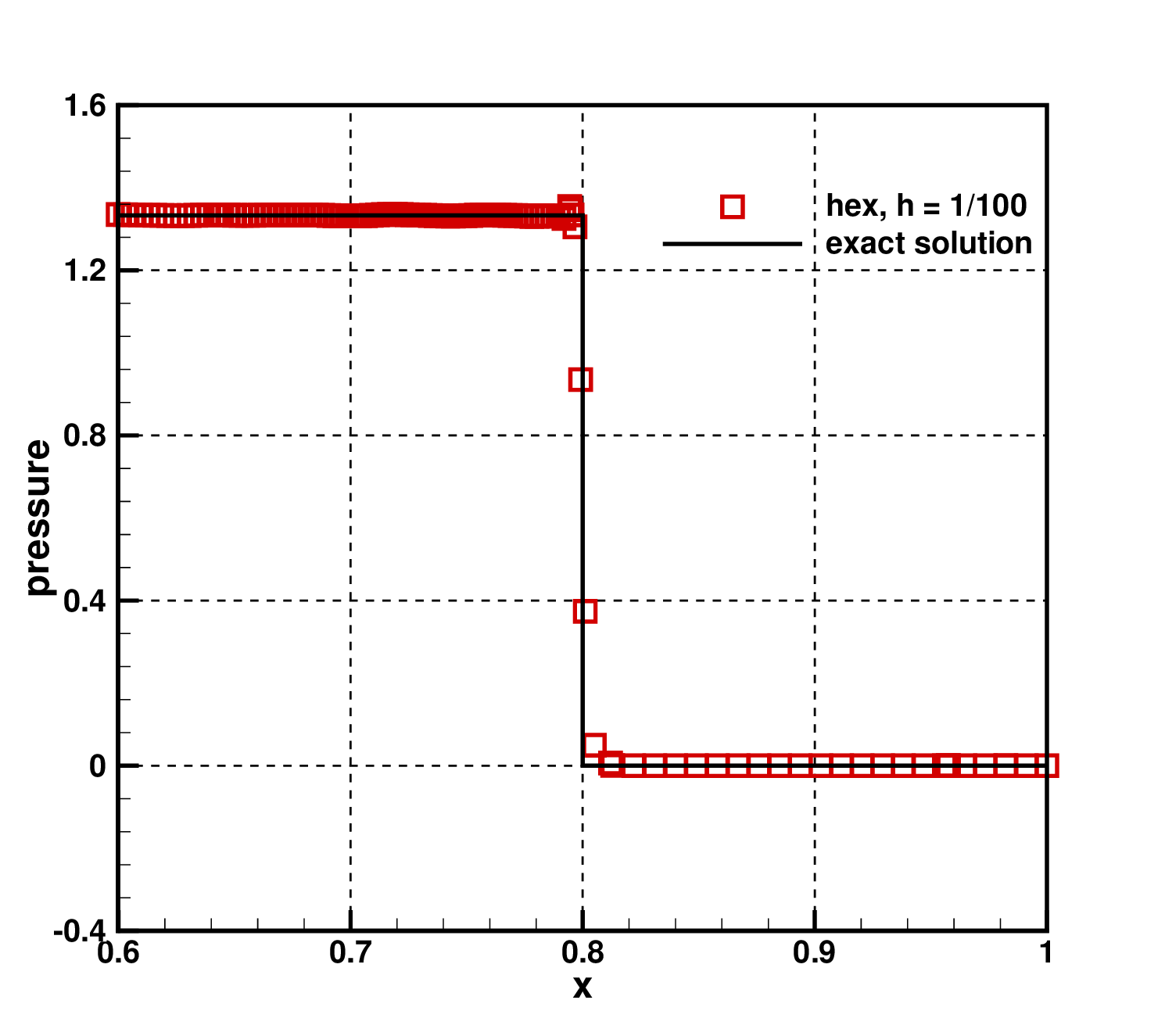}
\caption{\label{saltz-data-hex} Saltzman problem: the density and pressure profiles along the centerline with $h=1/100$ hexahedral mesh at $t=0.6$.}
\centering
\includegraphics[width=0.475\textwidth]{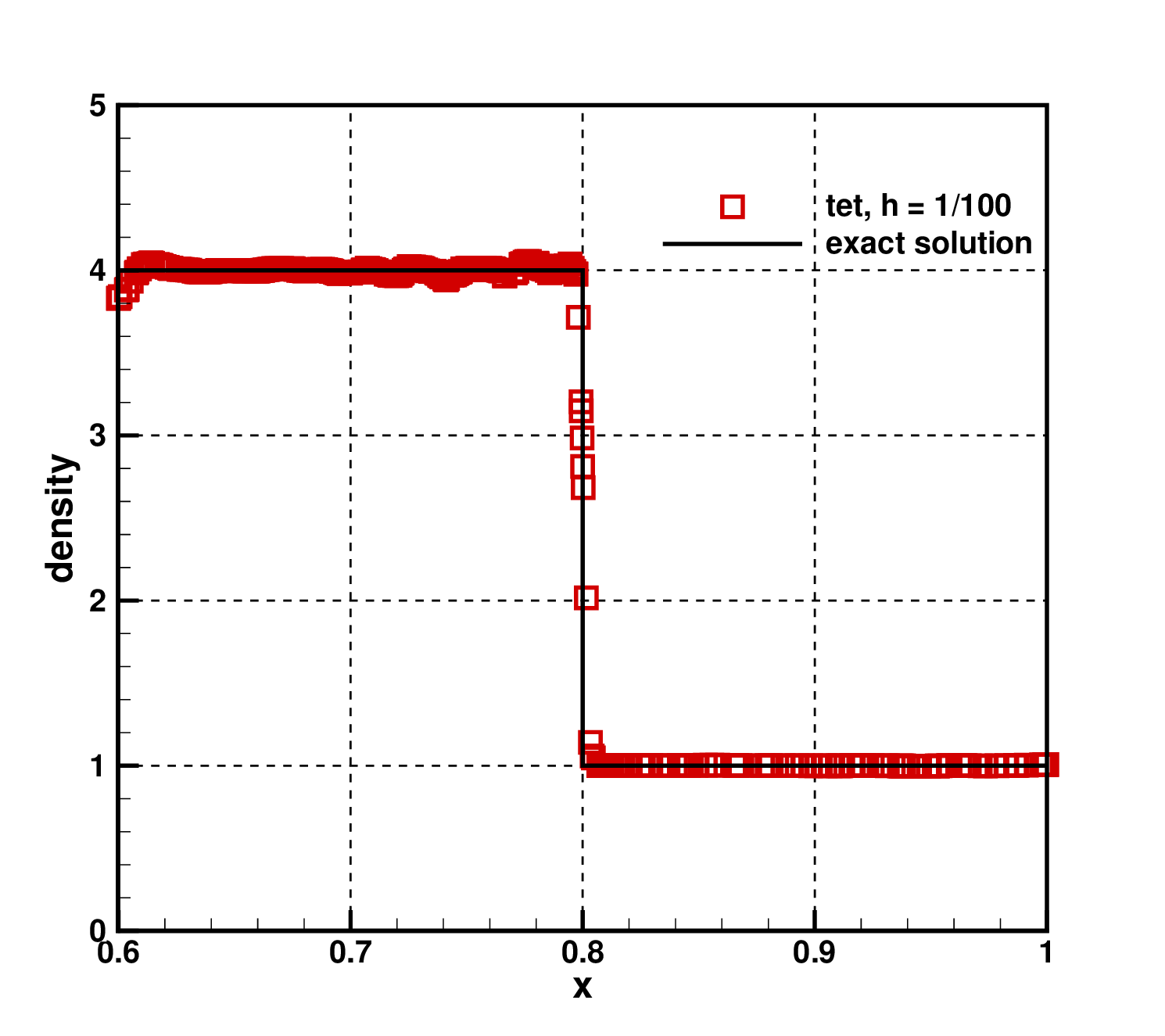} 
\includegraphics[width=0.475\textwidth]{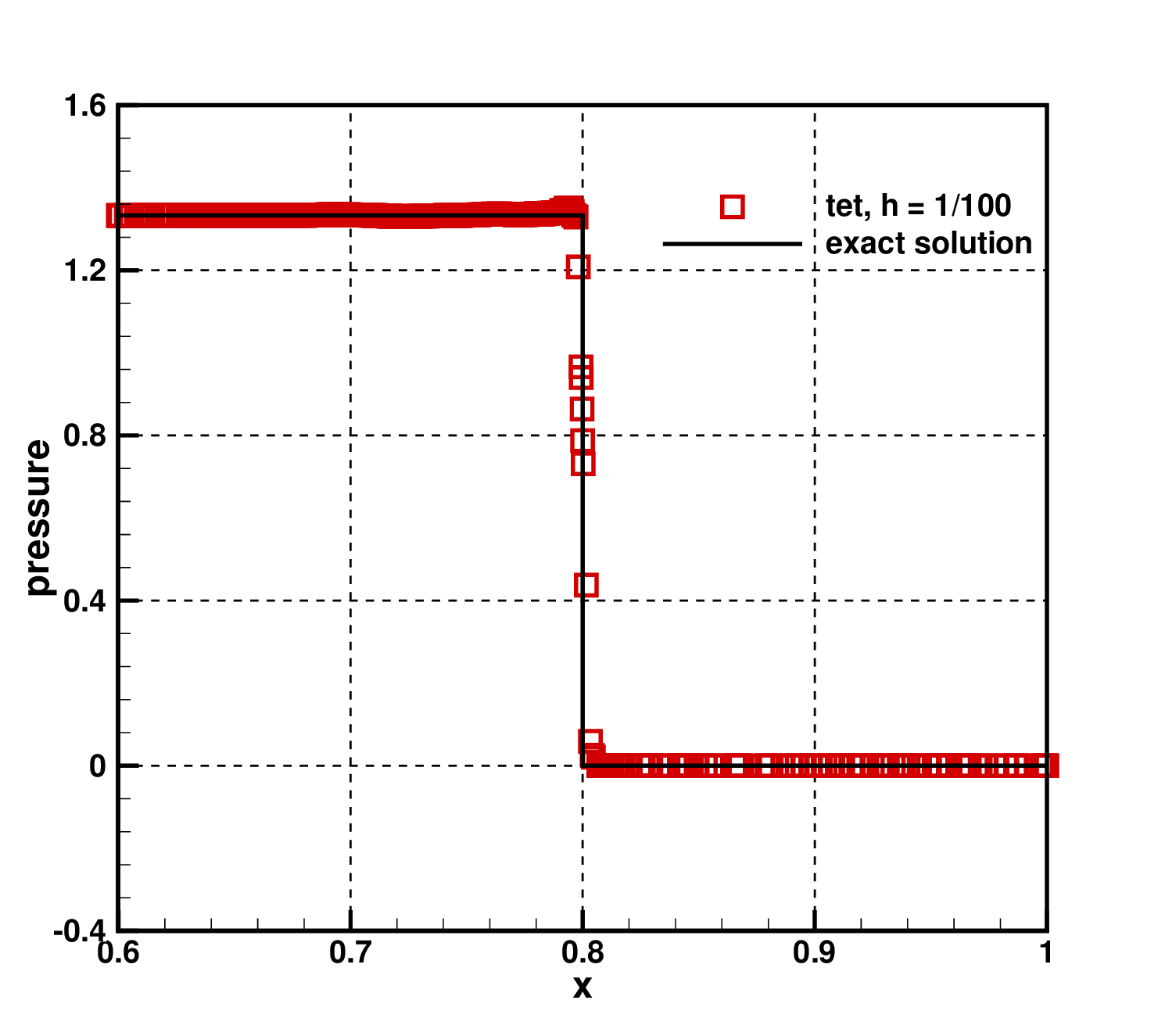}
\caption{\label{saltz-data-tet} Saltzman problem: the density and pressure profiles along the centerline with $h=1/100$ tetrahedral mesh at $t=0.6$.}
\end{figure}

\subsection{Saltzman problem}
The Saltzman problem describes the propagation of a planar shock wave, 
which is generated by the piston motion \cite{Lagrangian-case2,Lagrangian-case1}. 
During the shock propagation, the mesh is strongly compressed and distorted, 
which makes the Saltzman problem a benchmark case for validating the robustness of Lagrangian and ALE codes. 
The computational domain is $[0, 1]\times[0, 0.1]\times[0, 0.1]$. 
In this test, both hexahedral and tetrahedral meshes are used. 
The tetrahedral mesh is generated from the uniform mesh with cell size $h=1/100$.
Meanwhile, the hexahedral mesh is given by distorting a uniform mesh as
\begin{align*}
x_i &= \xi_i + (0.5\eta_i+\zeta_i-15\eta_i\zeta_i)\sin(\pi\xi_i),\\
y_i &= \eta_i,\\
z_i &= \zeta_i,
\end{align*}
where $(\xi_i,\eta_i,\zeta_i)$ denotes the original coordinates of vertex $i$ on the uniform $h=1/100$ hexahedral mesh. 
Initially, the box is filled with an ideal monatomic gas with $\rho=1, e=10^{-4}, \gamma=5/3$. 
The left-hand side wall acts as a piston with a constant velocity $U_{pis}=1$, 
and the other boundaries are reflective walls. 
For this case, the pure Lagrangian method will suffer from severe mesh distortion, 
while the ALE method can maintain good mesh quality through an appropriate mesh smoothing strategy. 
The Lagrangian velocity is chosen as nodal velocity 
and the smooth procedure is applied every 10 time steps for the hexahedral mesh and 5 time steps for the tetrahedral mesh. 
In the computation, 
a strong shock wave is generated from the left wall and propagates to the right faster than the piston. 
At $t=0.6$, the shock is expected to be located at $x=0.8$, 
and the post-hock solution is $\rho=4$ and $p=1.333$. 
The density and mesh distributions are presented in Fig.\ref{saltz-flow}. 
As can be seen, the post-shock meshes are severely distorted, especially for the tetrahedral mesh. 
The density and pressure profiles along the $y=z=0.05$ centerline are given 
in Fig.\ref{saltz-data-hex} and Fig.\ref{saltz-data-tet} for the hexahedral and tetrahedral meshes, respectively. 
The numerical profiles on both meshes agree well with the exact solution, 
capturing the correct shock position and post-shock states. 
Even on the severely distorted tetrahedral mesh, the shock structure is still well resolved.

\begin{figure}[!h]
\centering
\includegraphics[width=0.475\textwidth]{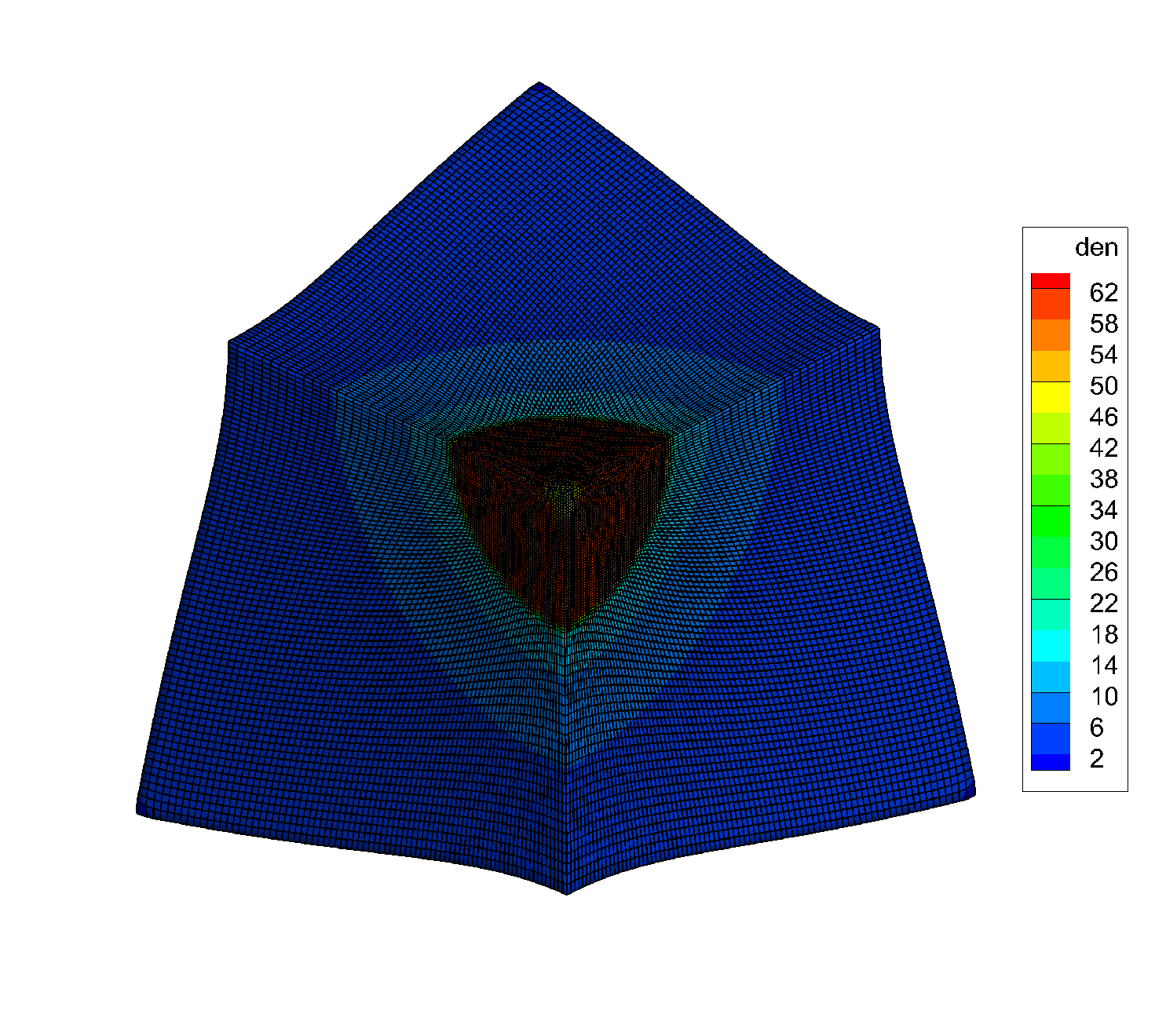}
\includegraphics[width=0.475\textwidth]{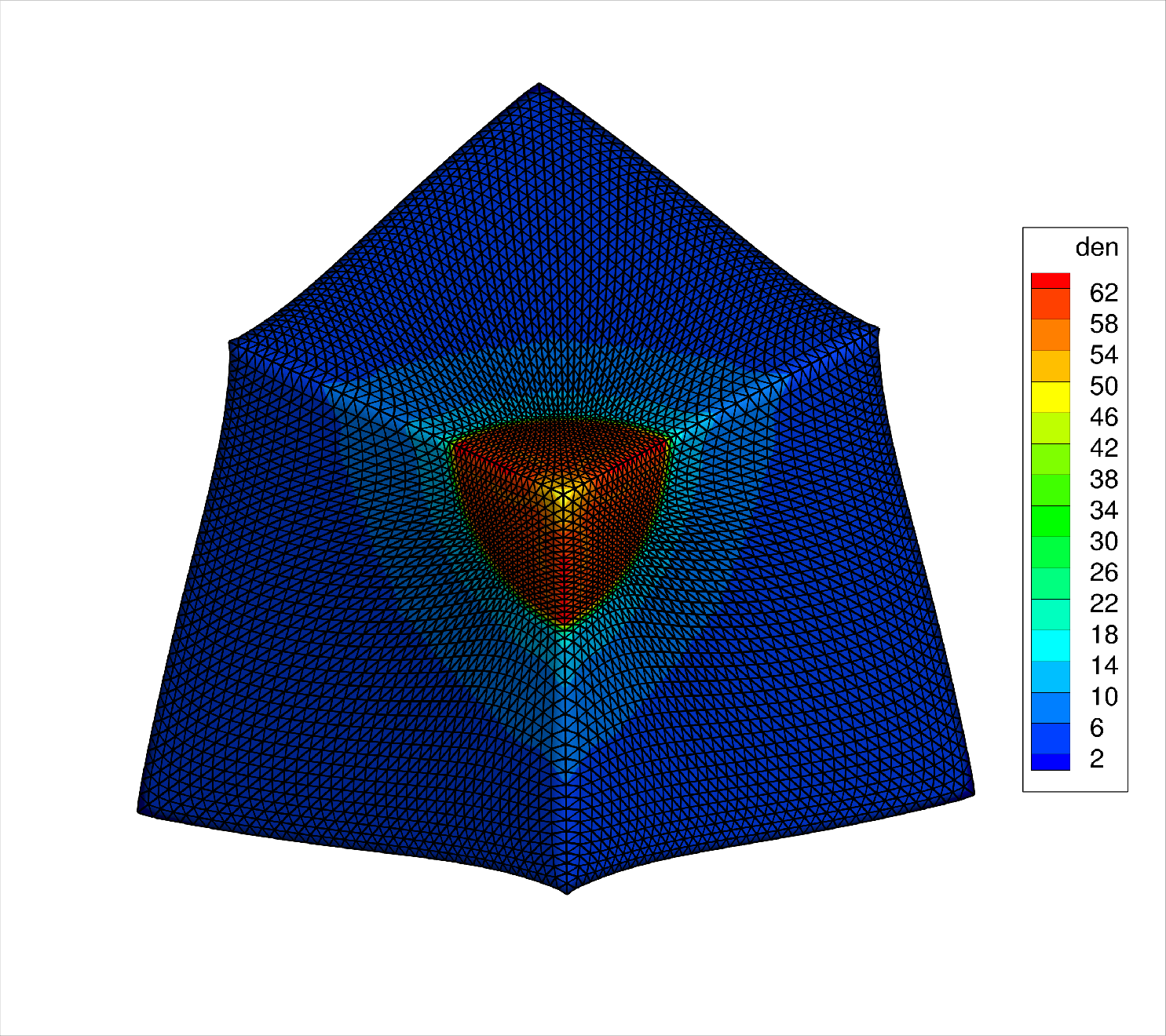}
\caption{\label{noh-flow} Noh problem: the mesh and density distributions at $t=0.6$ with $h=1/66$ hexahedral mesh (left) and $h=1/33$ tetrahedral mesh (right).}
\centering
\includegraphics[width=0.475\textwidth]{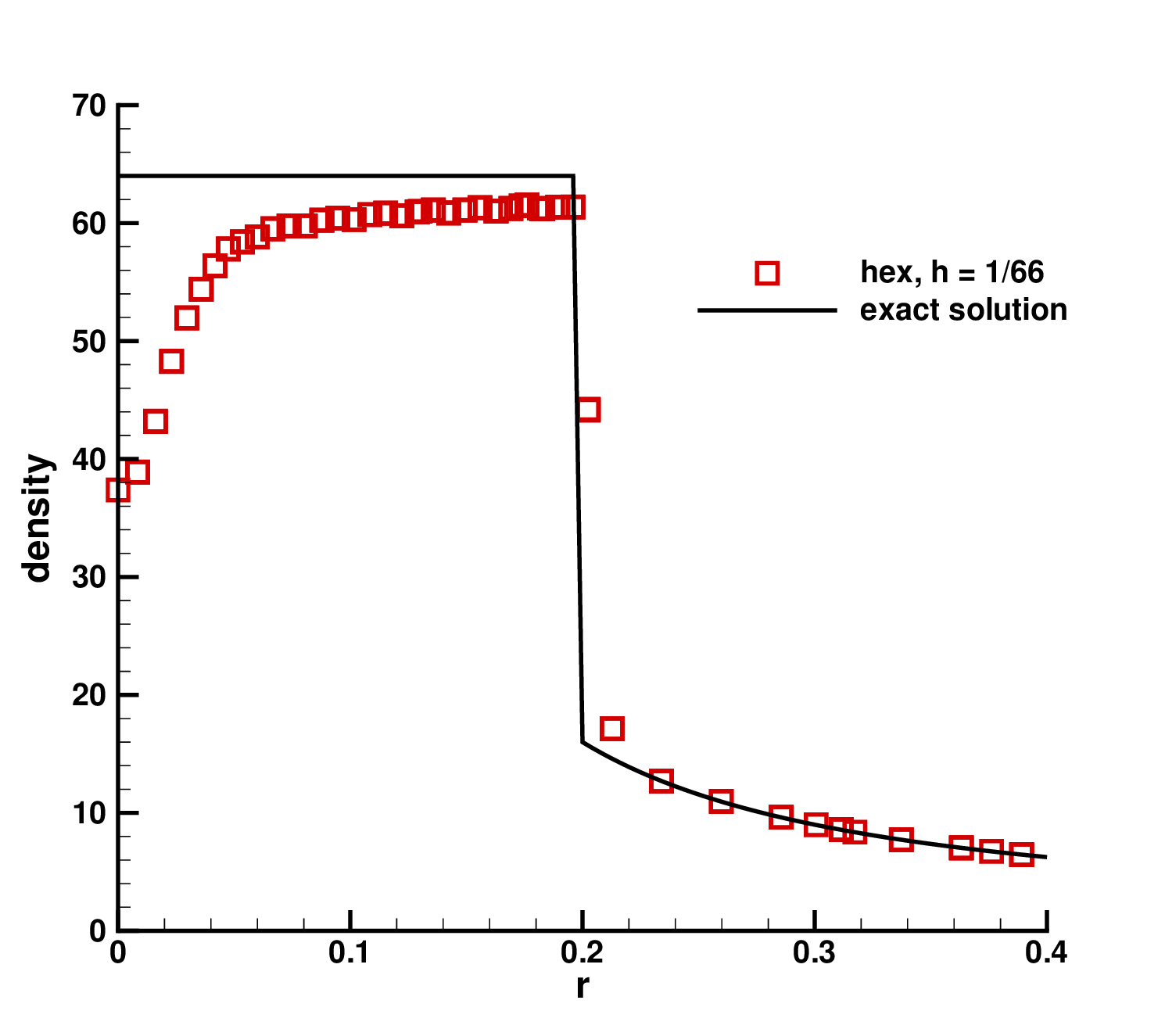}
\includegraphics[width=0.475\textwidth]{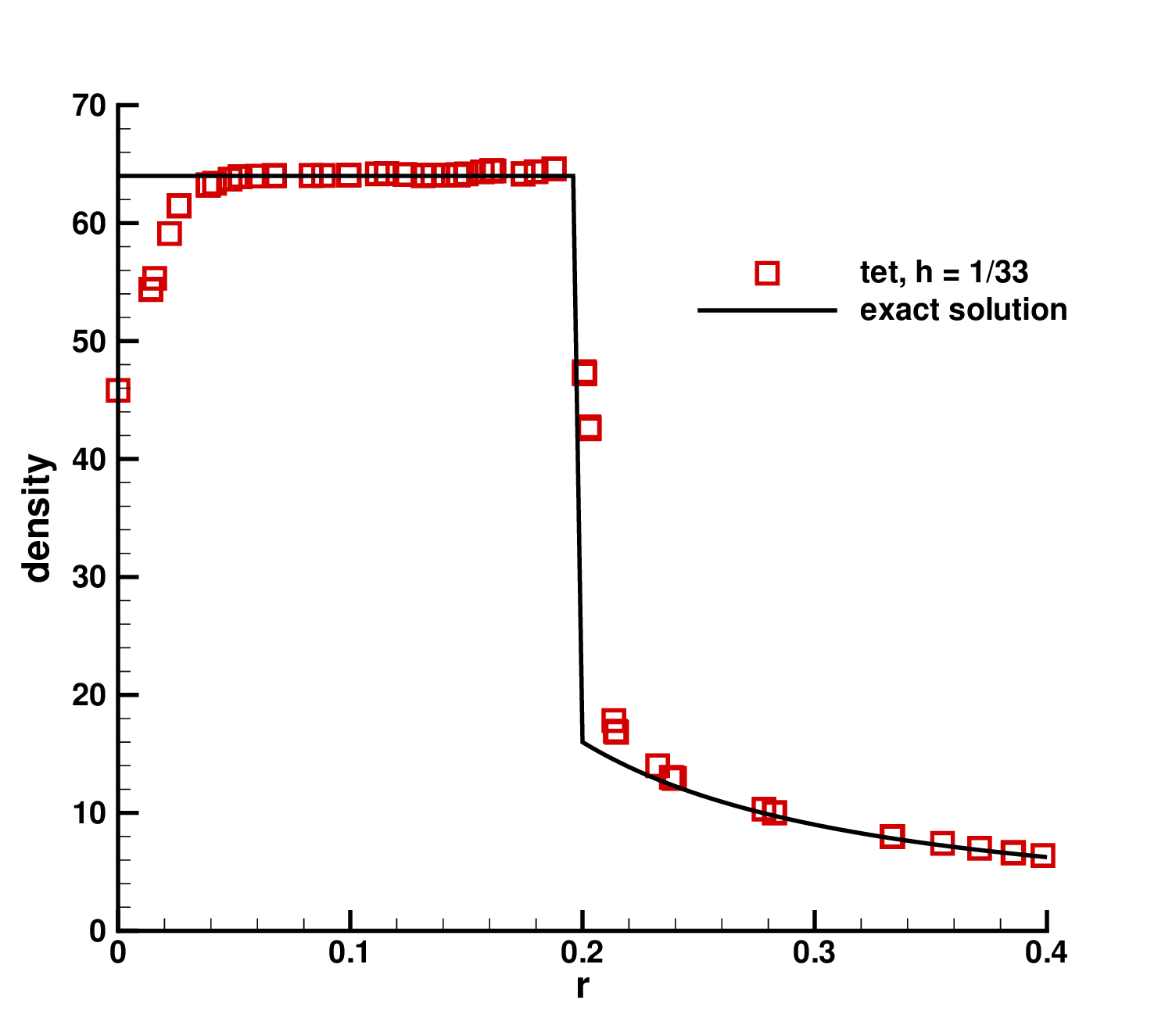}
\caption{\label{noh-data} Noh problem: the density profile along the diagonal line with $h=1/66$ hexahedral mesh (left) and $h=1/33$ tetrahedral mesh (right) at $t=0.6$.}
\end{figure}

\subsection{Noh problem}

The Noh problem models a highly compressible flow, 
where the gas with a uniform inward velocity impinges on the origin and generates a strong outward-moving shock wave.
This case is a three-dimensional extension of the two-dimensional Noh problem \cite{Case-Noh-1}, 
which is also a benchmark case for Lagrangian and ALE schemes. 
The initial computational domain is $[0,1.2]^3$. 
The monatomic gas is initialized with $\rho=1, e=10^{-4}$, 
and the initial velocity $\textbf{U}=(U,V,W)$ is given by
\begin{align*}
U=\frac{-x}{\sqrt{x^2+y^2+z^2}}, 
V=\frac{-y}{\sqrt{x^2+y^2+z^2}},
W=\frac{-z}{\sqrt{x^2+y^2+z^2}}.
\end{align*}
The reflective boundary condition is adopted on the symmetry planes containing the origin, 
and the other boundaries are free surfaces. 
In the computation, a spherical shock wave is generated at the origin and further propagates outward. 
At $t=0.6$, the shock is located at $r=0.2$,  and the exact solution for density is given by 
\begin{align*}
\rho=\begin{cases}
    64,        &r < 0.2,\\
    (1+t/r)^2, &r > 0.2. 
\end{cases}
\end{align*}
In the computation, the uniform hexahedral mesh with $h=1/66$ and the tetrahedral mesh $h=1/33$ are used initially. 
The Lagrangian velocity is adopted for mesh velocity,  and the smooth procedure is applied every 10 time steps. 
The mesh and density distributions at $t=0.6$ are presented in Fig.\ref{noh-flow} on the hexahedral and tetrahedral mesh. 
The density profile along the diagonal line is given in Fig.\ref{noh-data}.
As shown in Fig.\ref{noh-data}, the numerical solutions on both meshes agree well with the exact solution, 
which demonstrates that the present ALE method can robustly resolve the strong spherical shock even under highly compressed meshes.

\begin{figure}[!h]
\centering
\includegraphics[width=0.65\textwidth]{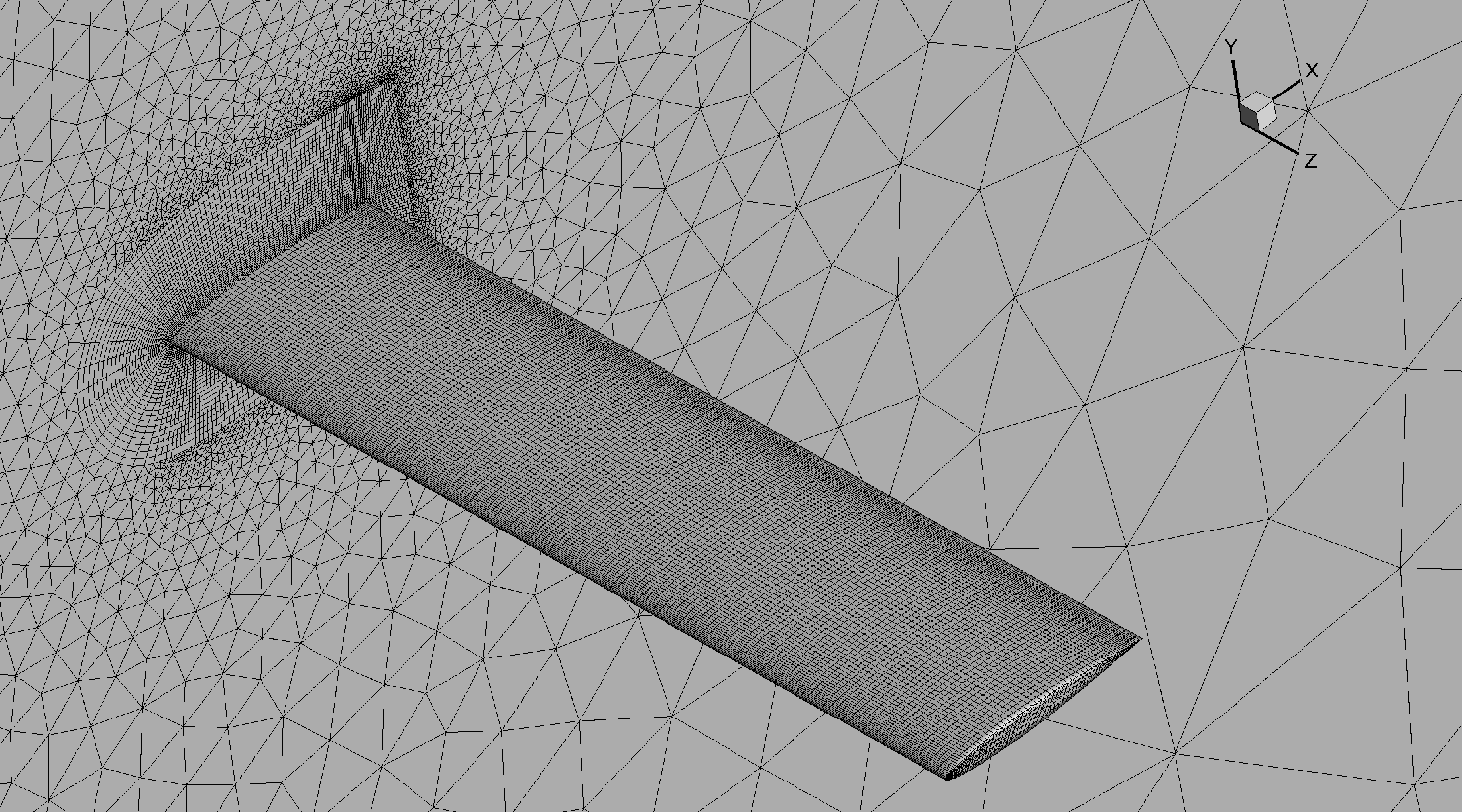}\\
\includegraphics[width=0.65\textwidth]{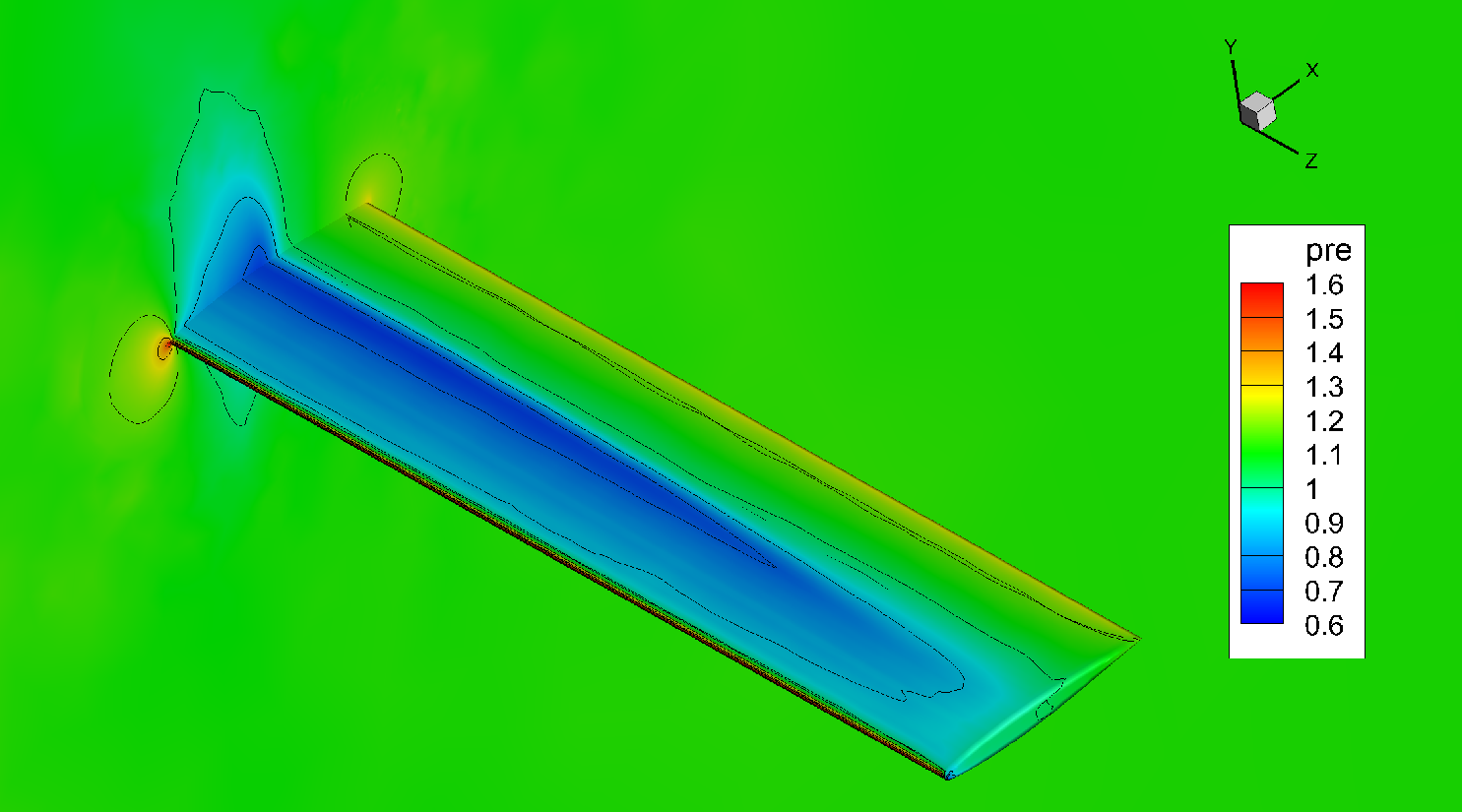}
\caption{\label{naca-flow} Flow over NACA 64A010 wing: the mesh (left) and pressure (right) distribution. }
\end{figure}

\begin{figure}[!h]
\centering
\includegraphics[width=0.475\textwidth]{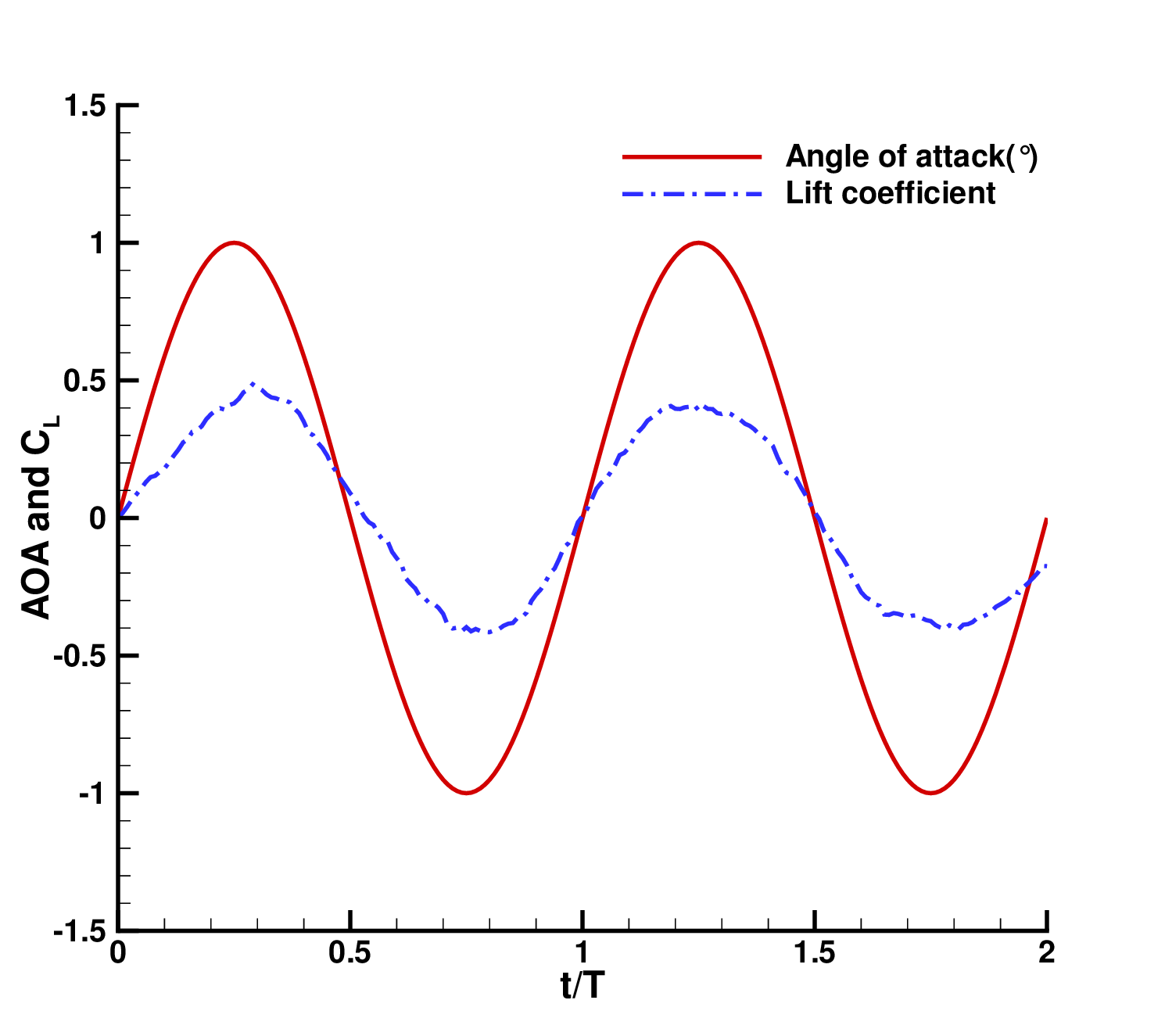}
\includegraphics[width=0.475\textwidth]{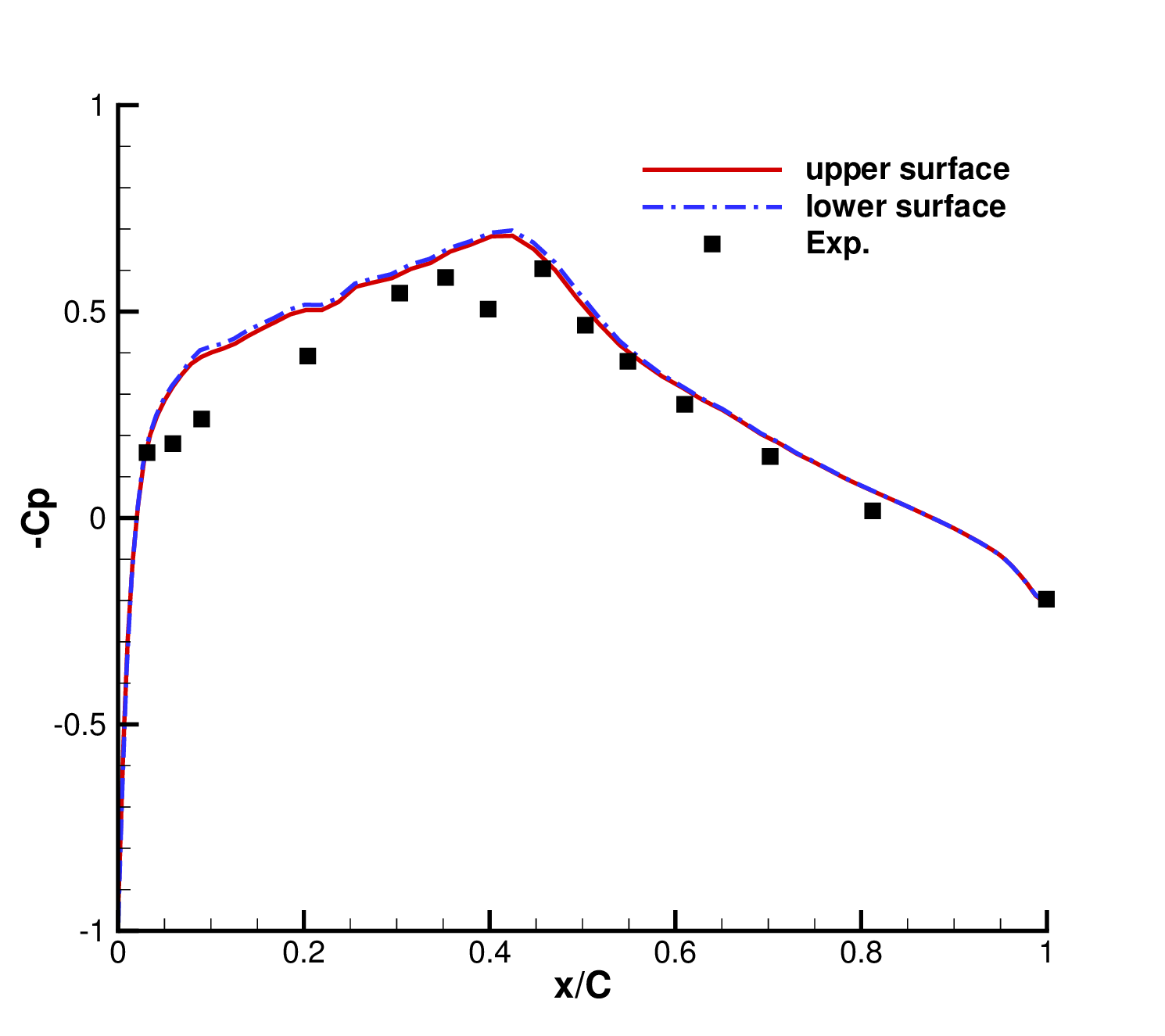} 
\caption{\label{naca-cl} Flow over NACA 64A010 wing: the time histories of $AOA$ and $C_L$ (left) and the mean pressure distribution on the wing at $z/b=0.5$ (right).}
\centering 
\includegraphics[width=0.475\textwidth]{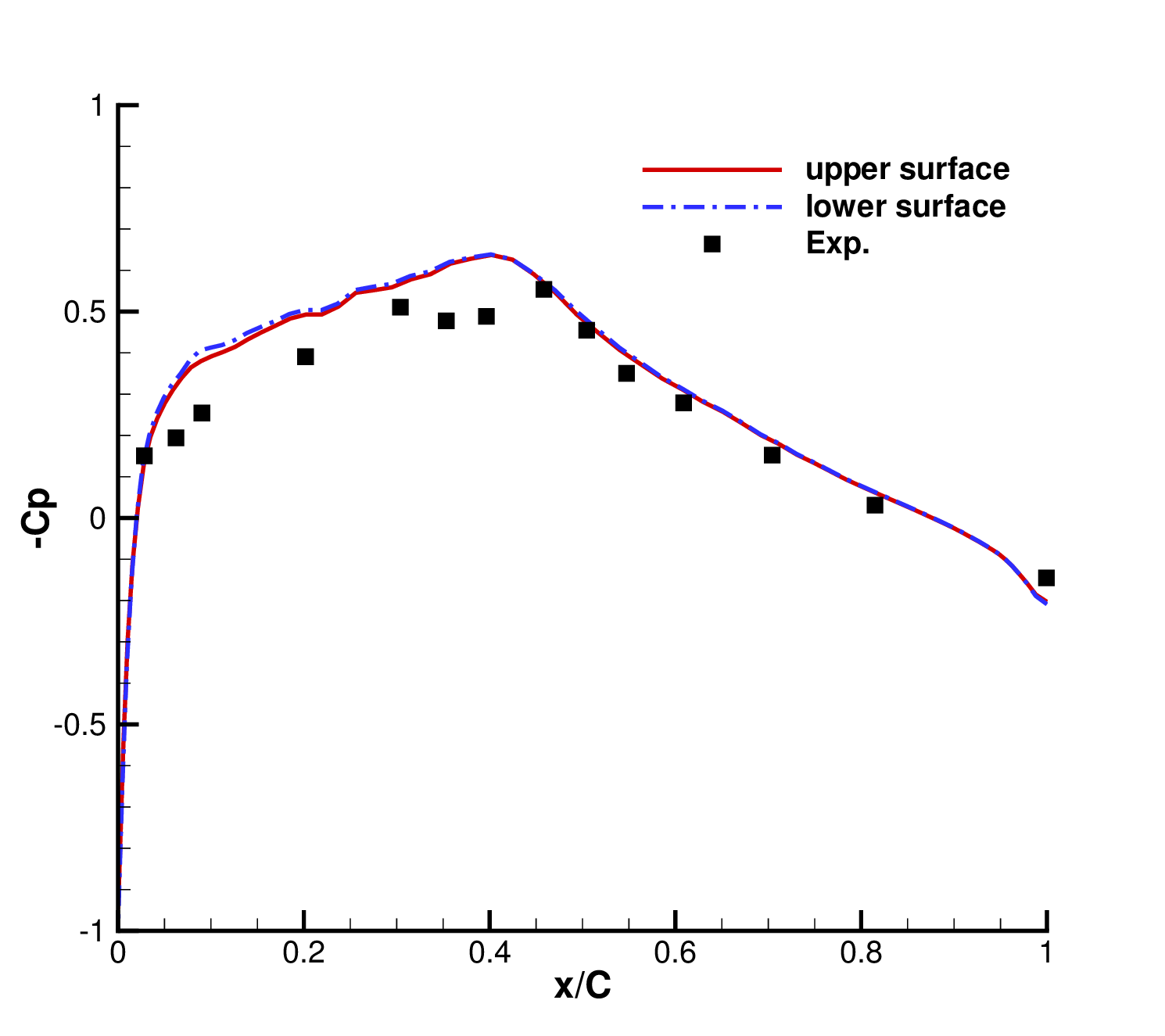}
\includegraphics[width=0.475\textwidth]{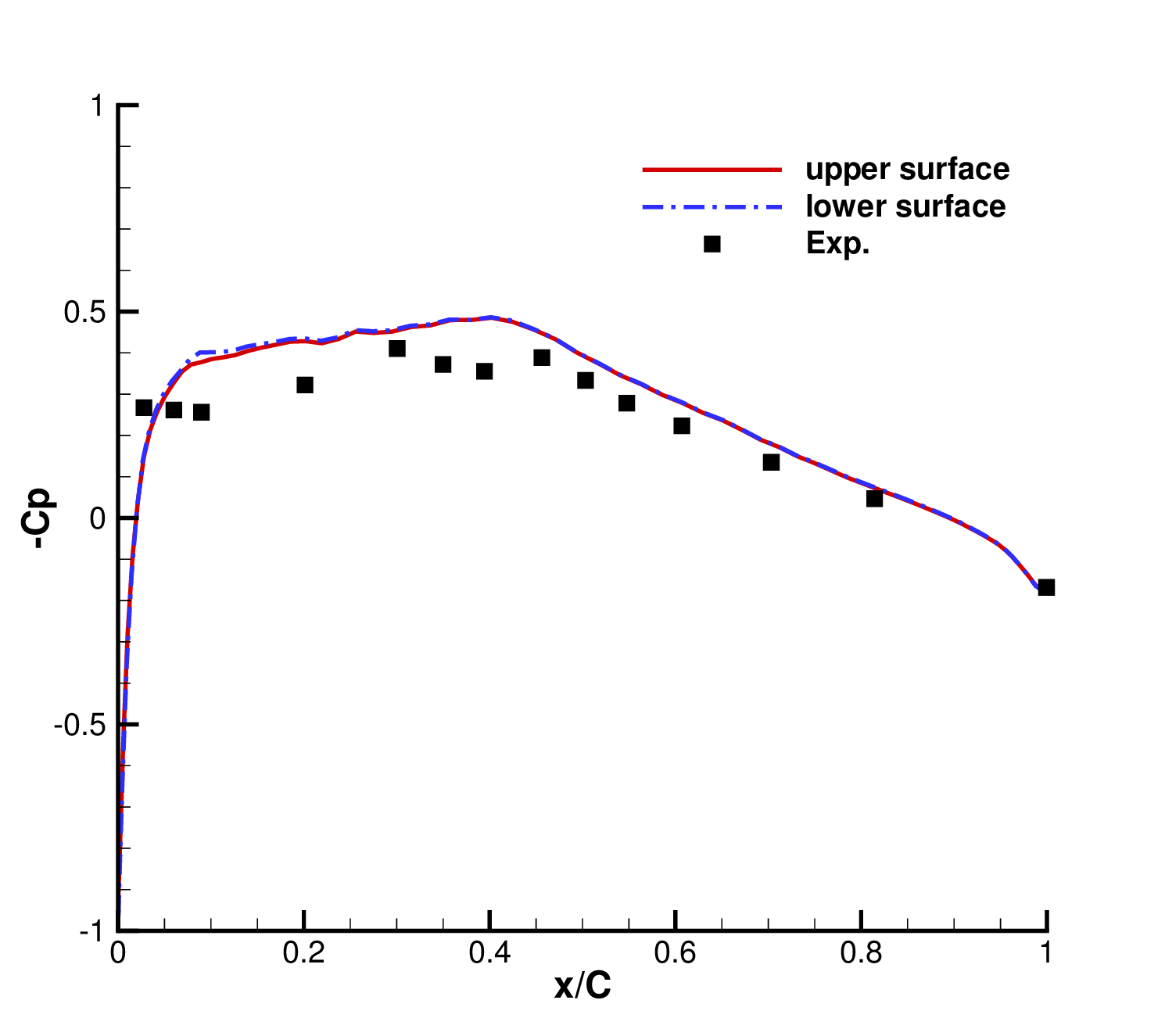}
\caption{\label{naca-cp2} Flow over NACA 64A010 wing: the mean pressure distributions on the wing at $z/b=0.766$ (left) and $z/b=0.94$ (right).}
\end{figure}

\subsection{Flow over NACA 64A010 wing}
In this case, the inviscid flow over an unswept pitching NACA 64A010 wing is presented to examine the 
performance of the current method for three-dimensional moving-boundary problems \cite{case-Naca64A010}. 
The wing undergoes a prescribed pitching motion about the mid-chord location ($50\%$ chord) as 
\begin{align*}
\alpha(t) = \alpha_0 + \alpha_m \sin(\omega t),
\end{align*}
where $\alpha_0 = 0^\circ$ is the mean angle of attack and $\alpha_m = 1^\circ$ is the pitching amplitude.  
The angular frequency $\omega$ is determined by 
\begin{align*}
k = \frac{\omega c}{2U_\infty},
\end{align*}
where $c=1$ is the chord length and $k=0.135$ is the reduced frequency. 
The computational domain is a spherical far-field with radius $25c$, 
and the grid consists of $1{,}722{,}478$ cells and $999{,}987$ nodes. 
Initially, the flow field is uniform with the freestream Mach number $Ma_\infty = 0.8$. 
The slip wall condition is adopted on the wing, 
and the symmetry boundary condition is used on the $z=0$ plane. 
In this computation, the mesh moves as a rigid body and follows the wing motion, 
and the entire computational grid rotates consistently with the wing without internal deformation.
The mesh and instantaneous pressure distribution are shown in Fig.\ref{naca-flow}, where the pressure contours exhibit clear spanwise variation. 
The time histories of the angle of attack $AOA$ and lift coefficient $C_L$ in the first two pitching periods are presented in Fig.\ref{naca-cl}. 
A distinct hysteresis behavior can be observed in the $C_L$-$AOA$ curve, 
reflecting the unsteady aerodynamic response induced by the prescribed pitching motion. 
The mean pressure coefficient distributions on the upper and lower surfaces of the wing in 
the second period are further compared at $z/b=0.5, 0.766, 0.94$ in Fig.\ref{naca-cl} and Fig.\ref{naca-cp2}, respectively. 
The computed pressure distributions agree well with the experimental data at both spanwise locations, 
showing that the present method can accurately predict the surface pressure response of the pitching wing.

\subsection{The helicopter rotor in hover state}

In this section, the Lynx rotor in the hover is simulated,  which is a configuration originally tested at 
NASA Ames Research Center's Outdoor Aerodynamic Research Facility. 
The rotor consists of four blades with NPL9615 airfoils and features a rectangular platform with zero twist. 
The detailed geometric specifications are documented in  \cite{Case-helicopter-1,Case-helicopter-2}. 
The chord length is $0.18 m$ and the radius is $R = 1.105 m$,  leading to an aspect ratio of $6.139$. 
The solidity of rotor is 0.208 with a blade root-cut at $0.384R$. 
The rotor operates with blade-tip Mach number $Ma_{tip} = 0.56$, 
pitching angle $\theta = 0^\circ$ and Reynolds number $Re = 2.31 \times 10^6$. 
The computational domain is set to $[10R,10R,10R]$, 
where $R$ is the rotor radius. 
The computational domain is discretized with a hybrid mesh, 
which contains approximately $37.56$ million cells. 
Hexahedral cells are employed in the vicinity of the rotor, 
and the minimum cell size is about $1.0 \times 10^{-3}\,\mathrm{m}$. 
Eight NVIDIA Tesla V100 GPUs are used for the parallel computation.

\begin{figure}[!h]
\centering
\includegraphics[width=0.495\textwidth]{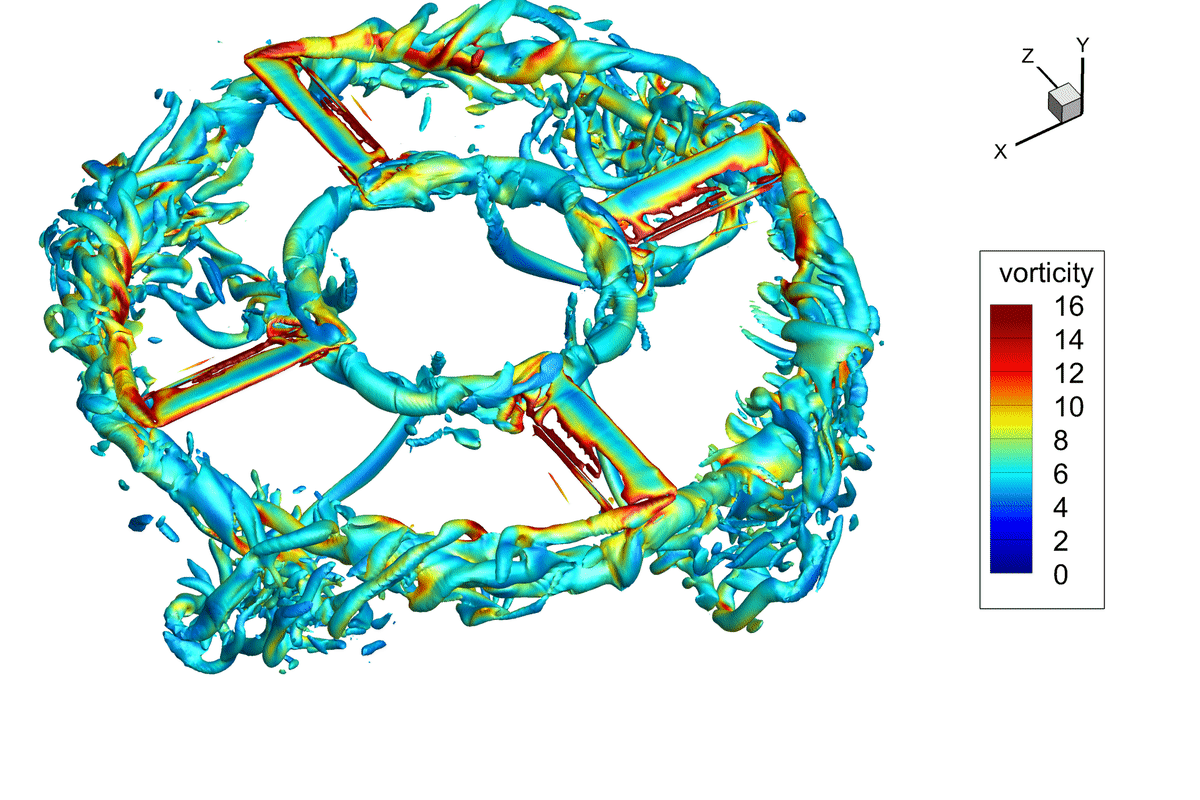}
\includegraphics[width=0.495\textwidth]{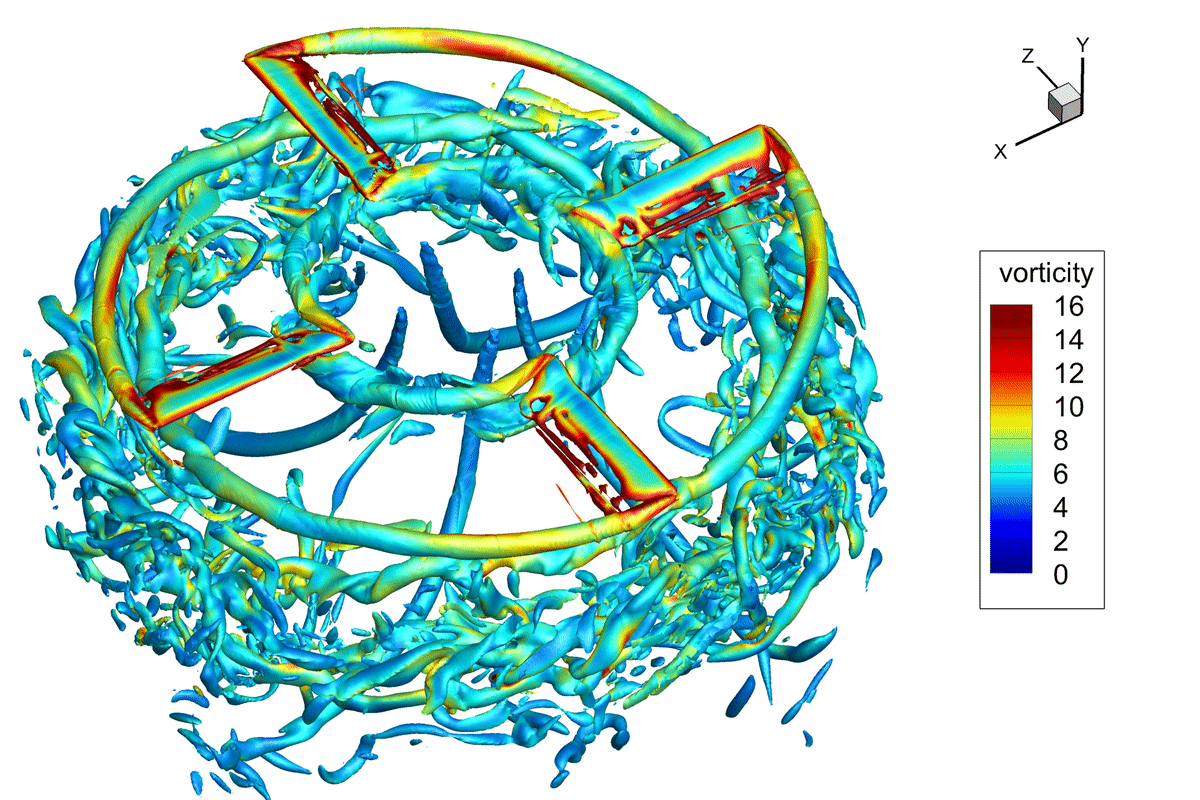}\\
\includegraphics[width=0.495\textwidth]{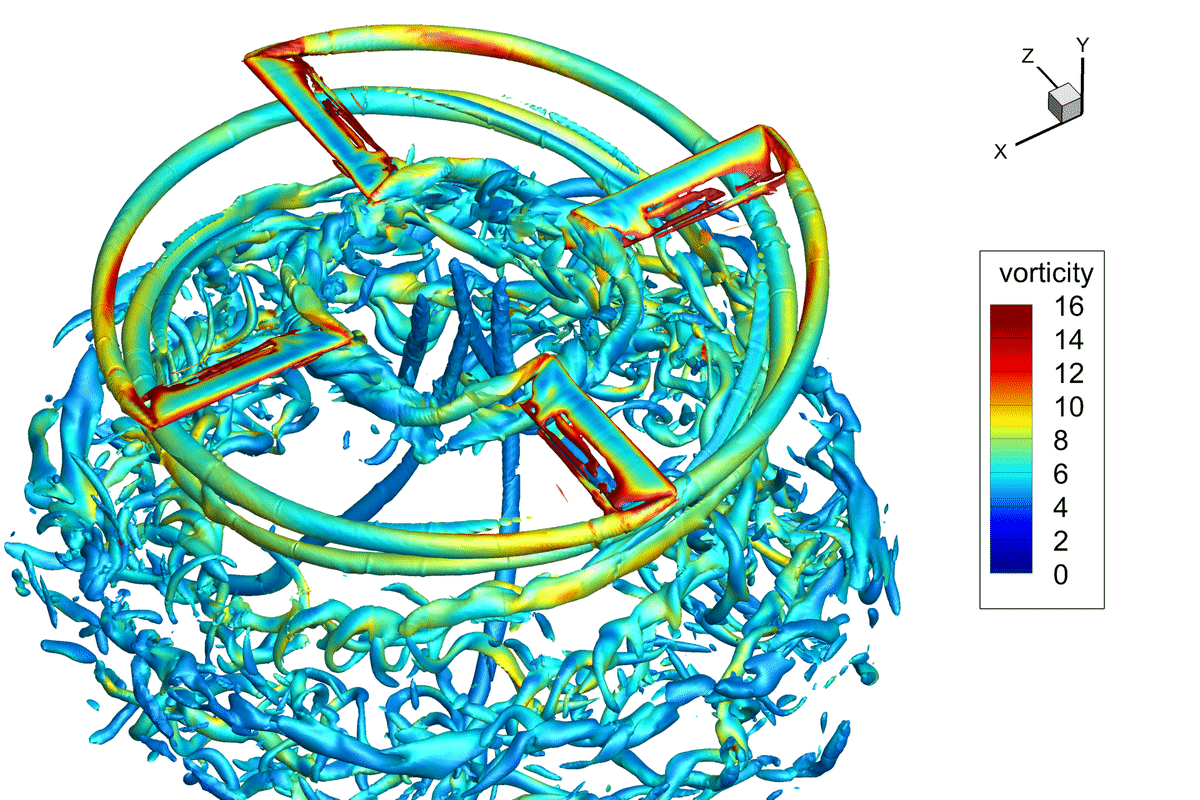}
\includegraphics[width=0.495\textwidth]{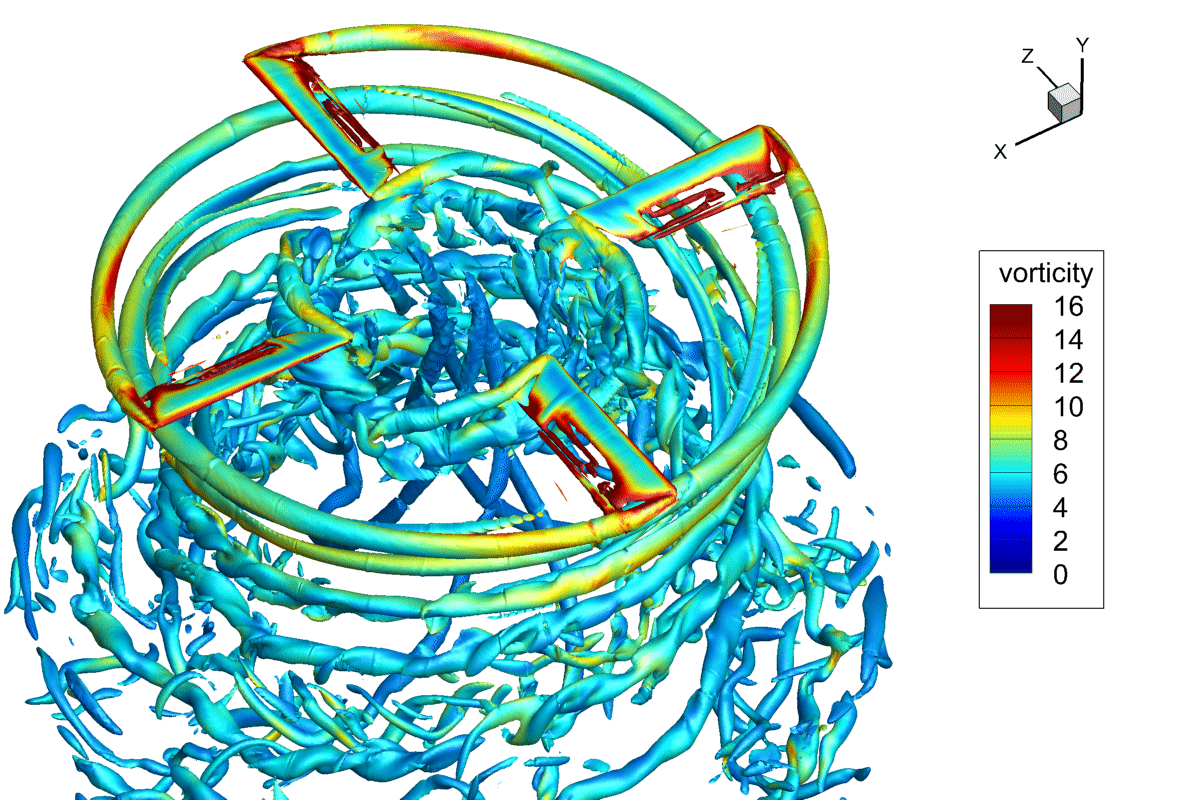}
\caption{\label{lynx-q} The helicopter rotor in hover state: the iso-surface of $Q=5$ colored by vorticity magnitude.}
\end{figure}

\begin{figure}[!h]
\centering
\includegraphics[width=0.45\textwidth]{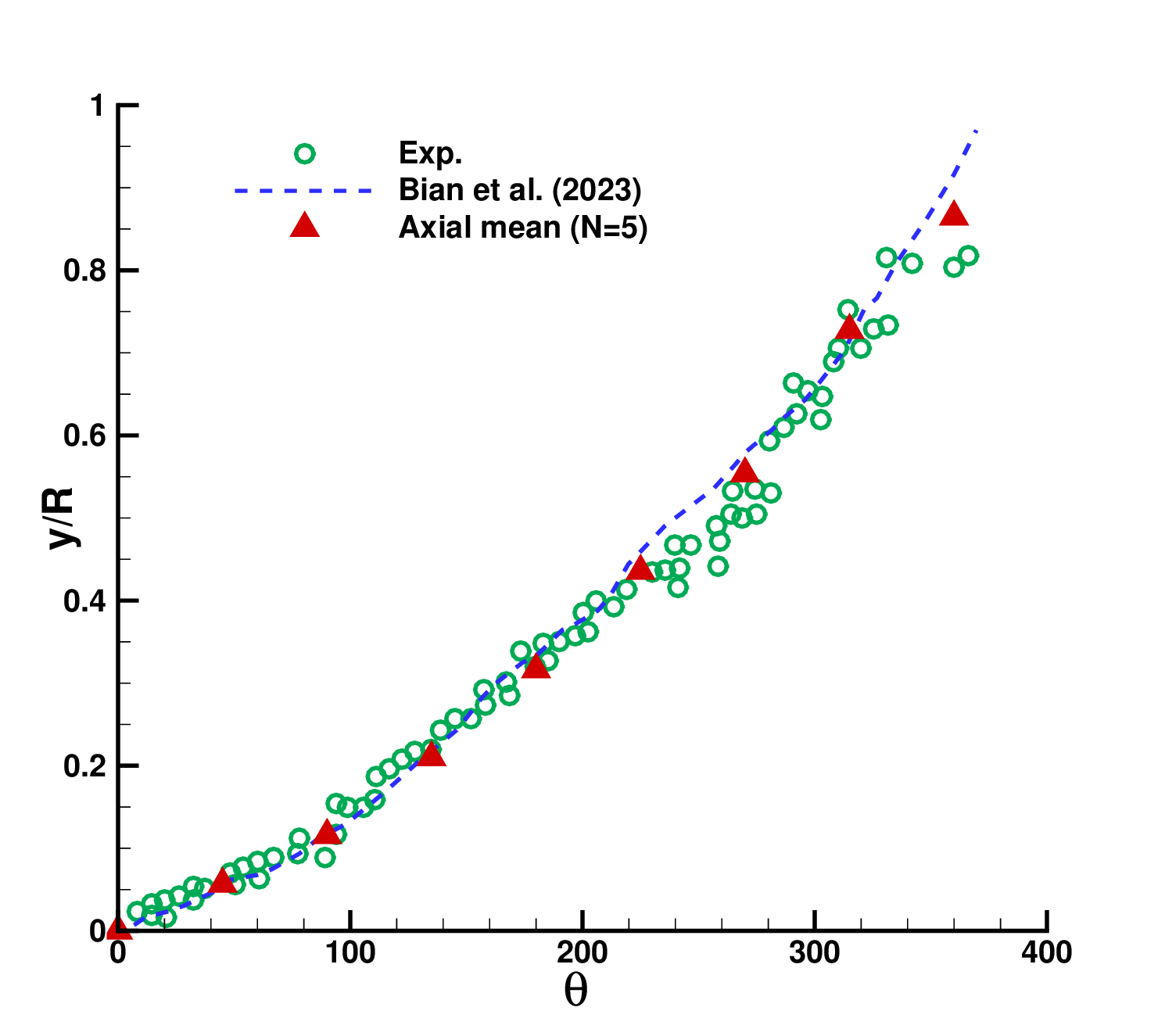}
\includegraphics[width=0.45\textwidth]{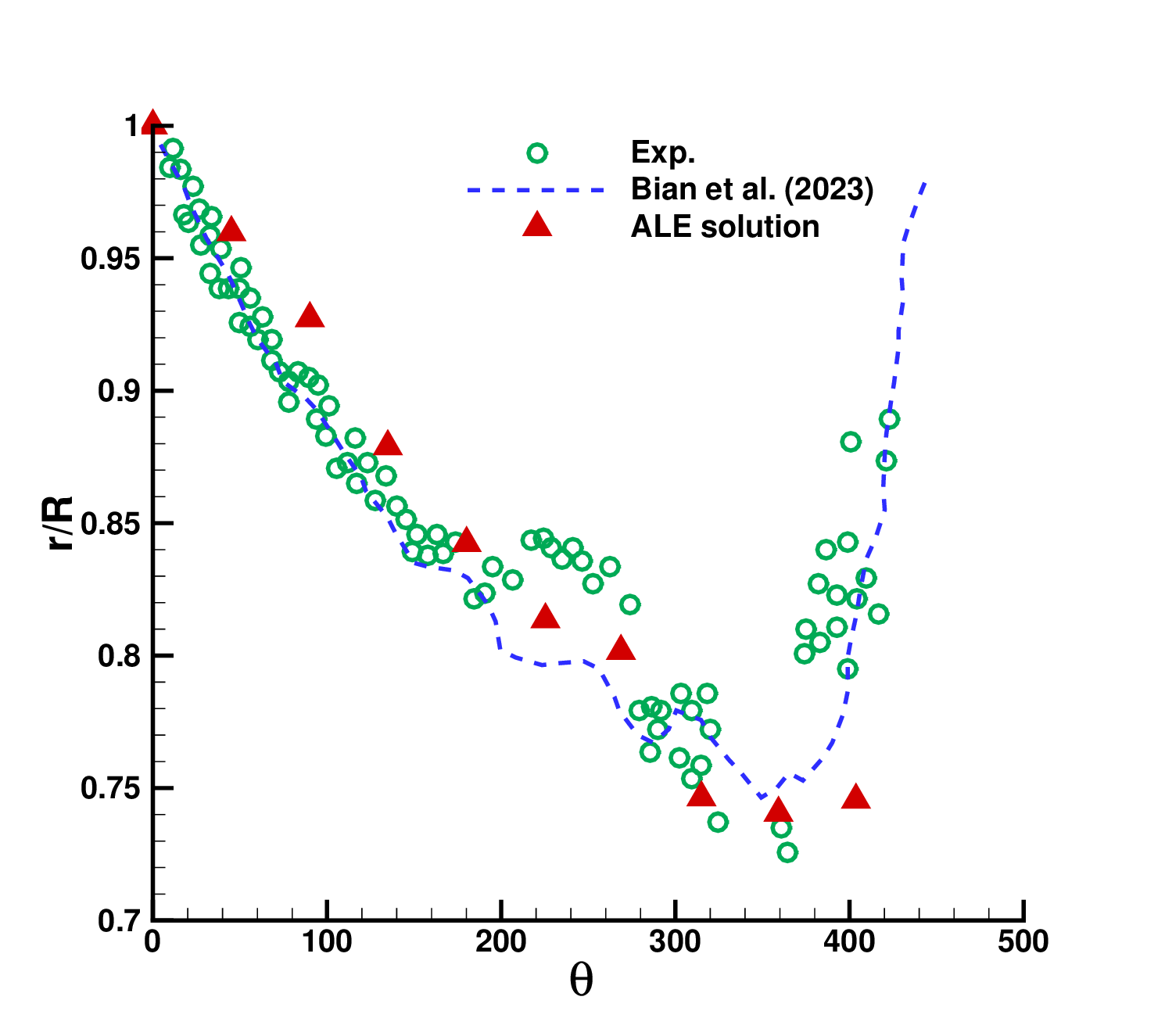}
\caption{\label{lynx-data} The helicopter rotor in hover state: 
the tip vortex trajectory of the Lynx rotor in axial direction (left) and radial direction (right).}
\end{figure}

Fig.\ref{lynx-q} presents the vortical structures of the hovering Lynx rotor, 
using the iso-surface of $Q=5$ colored by vorticity magnitude. 
The tip vortices shed from the rotating blades form distinct helical structures in the rotor wake, 
where there are numerous complex smaller-scale vortical structures as well. 
A quantitative comparison is provided in Fig.\ref{lynx-data}, 
which presents the axial and radial tip-vortex trajectories 
with experimental data \cite{Case-helicopter-1} and previous numerical results \cite{Case-helicopter-2}. 
$\theta$ denotes the wake age, 
and $y/R$ and $r/R$ represent the normalized axial and radial positions of the vortex core, respectively. 
The present results agree very well with both reference data, 
accurately reproducing the downstream convection and radial contraction of the tip vortices. 
These comparisons demonstrate the capability of the current scheme to capture the principal features of the rotor wake.

\section{Conclusion}

In this paper, a high-order compact ALE gas-kinetic scheme on three-dimensional hybrid meshes is developed for compressible flows with moving grids and moving boundaries. 
The new memory-reduction method is adopted for spatial discretization, 
which could construct a quadratic polynomial with only four least-square procedures sharing a single $3\times3$ coefficient matrix. 
To suppress numerical oscillations near discontinuities, 
the GENO reconstruction procedure is used to combine the quadratic polynomial $P^2(\boldsymbol{x})$ and a limited linear polynomial $\widetilde P^1(\boldsymbol{x})$. 
The two-stage fourth-order method is adopted for temporal discretization. 
To accelerate the computation, the current scheme is implemented on multiple GPUs using CUDA and MPI. 
Numerical examples are presented to validate the accuracy, efficiency and robustness of current scheme. 
Specifically, the current method is further applied to the pitching NACA 64A010 wing and the Lynx rotor in hover, 
demonstrating the applicability of the scheme to three-dimensional moving-boundary problems. 
These results validate the performance of the current compact ALE scheme for large-scale three-dimensional moving-mesh and moving-boundary computations. 
In the future,  the method will be extended to problems involving sliding meshes and fluid-structure interaction (FSI).

\section*{Acknowledgements}
The current research of L. Pan is supported by National Natural Science Foundation of China (12471364, 12494543),  
National Numerical Wind Tunnel Project, Beijing Natural Science Foundation (L2608005) and Science Challenge Project (No. TZ2025001).  
The current research of X. Ji is supported by the Funding of National Key Laboratory of Computational Physics, 
the National Natural Science Foundation of China (No 12302378), the National Numerical Wind Tunnel Project.

\section*{Declaration of competing interest}
The authors declare that they have no known competing financial interests or personal relationships 
that could have appeared to influence the work reported in this paper.

\section*{Data availability}
The data that support the findings of this study are available from
the corresponding author upon reasonable request.


\begin{thebibliography}{10}

\bibitem{ALE-2} A.J. Barlow, P.H. Maire, W.J. Rider, R.N. Rieben, M.J. Shashkov, Arbitrary Lagrangian-Eulerian methods for modeling high-speed compressible multimaterial flows, J. Comput. Phys. 322 (2016) 603-665.

\bibitem{Lagrangian-Benson} D.J. Benson, Computational methods in Lagrangian and Eulerian hydrocodes, Comput. Methods Appl. Mech. Eng. 99 (1992) 235-394.

\bibitem{BGK-1} P.L. Bhatnagar, E.P. Gross, M. Krook, A Model for Collision Processes in Gases I: Small Amplitude Processes in Charged and Neutral One-Component Systems, Phys. Rev. 94 (1954) 511-525.

\bibitem{Case-helicopter-2} W. Bian, G. Zhao, X. Chen, B. Wang, Q. Zhao, High-fidelity simulation of blade vortex interaction of helicopter rotor based upon TENO scheme, Chin. J. Aeronaut. 36 (2023) 275-292.

\bibitem{ALE-direct-1} W. Boscheri, High order direct Arbitrary-Lagrangian-Eulerian (ALE) finite volume schemes for hyperbolic systems on unstructured meshes, Arch. Comput. Methods Eng. 24 (2017) 1–51.

\bibitem{ALE-direct-2} W. Boscheri, M. Dumbser, High order accurate direct Arbitrary-Lagrangian-Eulerian ADER-WENO finite volume schemes on moving curvilinear unstructured meshes, Computers and Fluids 136 (2016) 48-66.

\bibitem{Lagrangian-case2} E.J. Caramana, C.L. Rousculp, D.E. Burton, A compatible, energy and symmetry preserving Lagrangian hydrodynamics algorithm in three-dimensional Cartesian geometry, J. Comput. Phys. 157 (2000) 89-119.

\bibitem{BGK-2} S. Chapman, T.G. Cowling, The Mathematical theory of Non-Uniform Gases, third edition, Cambridge University Press, (1990).

\bibitem{GRP-high} Z.F. Du, J.Q. Li, A Hermite WENO reconstruction for fourth order temporal accurate schemes based on the GRP solver for hyperbolic conservation laws, J. Comput. Phys. 355 (2018) 385-396.

\bibitem{Lagrangian-case1} J.K. Dukowicz, B.Meltz, Vorticity errors in multidimensional lagrangian codes. J. Comput. Phys. 99 (1992) 115-134.

\bibitem{case-Naca64A010} A.L. Gaitonde, A dual-time method for the solution of the unsteady Euler equations, Aeronautical Journal October 1994.

\bibitem{ALE-1} C.W. Hirt, A.A. Armsden, J.L. Cook, An arbitrary Lagrangian Eulerian computing method for all flow speed, J. Comput. Phys. 135 (1997) 203-216.

\bibitem{GKS-compact-gcf} X. Ji, W. Shyy, K. Xu, A gradient compression-based compact high-order gas-kinetic scheme on 3D hybrid unstructured meshes, Int. J. Comput. Fluid D 35 (2021) 485509.

\bibitem{GKS-ALE1} C.Q. Jin, K. Xu, A unified moving grid gas-kinetic method in Eulerian space for viscous flow computation, J. Comput. Phys. 222 (2007) 155-175.

\bibitem{Case-Sedov} J.R. Kamm, F.X. Timmes, On efficient generation of numerically robust Sedov solutions, Technical Report LA-UR-07-2849, Los Alamos National Laboratory, 2007.

\bibitem{METIS-Kar} G. Karypis, V. Kumar, METIS: a software package for partitioning unstructured graphs, partitioning meshes, and computing fill-reducing orderings of sparse matrices version 5.1.0, http://www.cs.umn.edu/karypis, 2013.

\bibitem{ALE-rezone} P. Knupp, L.G. Margolin, M. Shashkov, Reference Jacobian optimization-based rezone strategies for arbitrary Lagrangian Eulerian methods, J. Comput. Phys. 176 (2002) 93-128.

\bibitem{ALE-remap} M. Kucharik, M. Shashkov, Conservative multi-material remap for staggered multi-material Arbitrary Lagrangian-Eulerian methods, J. Comput. Phys. 258 (2014) 268-304.

\bibitem{Case-helicopter-1} J.S. Light, Tip vortex geometry of a hovering helicopter rotor in ground eﬀect, J. Am. Helicopter Soc. 38 (1993) 34-42.

\bibitem{GKS-MR-compact} H.Y. Liu, X. Ji, Y.P. Mao, Z. Qian, K. Xu, A Memory Reduction Compact Gas Kinetic Scheme on 3D Unstructured Meshes, Comput. Phys. Commun. 320 (2026) 109981.

\bibitem{Lagrangian-c1} P.H. Maire, R. Abgrall, J. Breil, J. Ovadia, A cell-centered Lagrangian scheme for two-dimensional compressible flow problems, SIAM J. Sci. Comput. 29 (2007) 1781-1824.

\bibitem{IBM-direct-Meng} B.Q. Meng, J.S. Zeng, S. Li, B.L. Tian, J.H. Liu, A particle-resolved direct numerical simulation method for the compressible gas flow and arbitrary shape solid moving with a uniform framework, Comput. Phys. Commun. 303 (2024) 109266.

\bibitem{IBM-Mittal} R. Mittal, G. Iaccarino, Immersed boundary methods, Annu. Rev. Fluid Mech. 37 (2005) 239-261.

\bibitem{GKS-ALE2} G.X. Ni, S. Jiang, K. Xu, Remapping-free ALE-type kinetic method for flow computations, J. Comput. Phys. 228 (2009) 3154-3171.

\bibitem{Case-Noh-1} W.F. Noh, Errors for calculations of strong shocks using artificial viscosity and an artificial heat flux, J. Comput. Phys. 72 (1987) 78-120.

\bibitem{GKS-high-1} L. Pan, K. Xu, Q. Li, J. Li, An efficient and accurate two-stage fourth-order gas-kinetic scheme for the Navier-Stokes equations, J. Comput. Phys. 326 (2016) 197-221.

\bibitem{GKS-high-2} L. Pan, J. Li, K. Xu, A few benchmark test cases for higher-order Euler solvers, Numer. Math., Theory Methods Appl. 10 (2017) 711-736.

\bibitem{GKS-ALE4} L. Pan, F.X. Zhao, K. Xu, High-order ALE gas-kinetic scheme with WENO reconstruction, J. Comput. Phys. 417 (2020) 109558.

\bibitem{GKS-ALE5} L. Pan, K. Xu, An Arbitrary-Lagrangian-Eulerian High-Order Gas-Kinetic Scheme for Three-Dimensional Computations, J. Sci. Comput. (2021) 88:8.

\bibitem{GCL-2} P.O. Persson, J. Bonet, J. Peraire, Discontinuous Galerkin solution of the Navier-Stokes equations on deformable domains, Comput. Methods Appl. Mech. Eng. 198 (2009) 1585-1595.

\bibitem{GKS-ALE3} X.D. Ren, K. Xu, W. Shyy, A multi-dimensional high-order DG-ALE method based on gas-kinetic theory with application to oscillating bodies, J. Comput. Phys. 316 (2016) 700-720.

\bibitem{Overset-Steger} J.L. Steger, J.A. Benek, On the use of composite grid schemes in computational aerodynamics, Comput. Methods Appl. Mech. Eng. 64 (1987) 301-320.

\bibitem{GCL-1} P. Thomas, C. Lombard, Geometric conservation law and its application to flow computations on moving grids, AIAA J. 17 (1979) 1030-1037.

\bibitem{IBM-Uhlmann} M. Uhlmann, An immersed boundary method with direct forcing for the simulation of particulate flows, J. Comput. Phys. 209 (2005) 448-476.

\bibitem{GKS-ALE6} Y.B. Wang, X. Ji, L. Pan, An efficient and robust high-order compact ALE gas-kinetic scheme, J. Comput. Phys. 543 (2025) 114398

\bibitem{GKS-GPU-str-2023} Wang YH, Cao GY, Pan L. Multiple-GPU accelerated high-order gas-kinetic scheme for direct numerical simulation of compressible turbulence, J. Comput. Phys. 476 (2023) 111899.

\bibitem{GKS-GPU-uns-2025} Wang YH, Cao WX, Pan L. Multiple-GPU accelerated high-order gas-kinetic scheme on three-dimensional unstructured meshes, Comput. Phys. Commun. 310 (2025) 109513.

\bibitem{GKS-Xu1} K. Xu, A gas-kinetic BGK scheme for the Navier-Stokes equations and its connection with artificial dissipation and Godunov method, J. Comput. Phys. 171 (2001) 289-335.

\bibitem{GKS-Xu2} K. Xu, Direct modeling for computational fluid dynamics: construction and application of unified gas kinetic schemes, World Scientific (2015).

\bibitem{GKS-noncompact-1} Y.Q Yang, L.Pan, K. Xu, High-order gas-kinetic scheme on three-dimensional unstructured meshes for compressible flows, Phys. Fluids 33 (2021) 096102.

\bibitem{GKS-noncompact-2} Y.Q. Yang, L. Pan, K. Xu, Three-dimensional third-order gas-kinetic scheme on hybrid unstructured meshes for Euler and Navier-Stokes equations, Comput. Fluids 255 (2023) 105834.

\bibitem{IBM-cutcell} T. Ye, R. Mittal, H.S. Udaykumar, W. Shyy, An accurate Cartesian grid method for viscous incompressible flows with complex immersed boundaries, J. Comput. Phys. 156 (1999) 209-240.

\bibitem{WENO-unstructured} F.X. Zhao, L. Pan, S.H. Wang, Weighted essentially non-oscillatory scheme on unstructured quadrilateral and triangular meshes for hyperbolic conservation laws, J. Comput. Phys. 374 (2018) 605-624.

\bibitem{GKS-GENO} F.X. Zhao, K. Xu, A generalized ENO reconstruction in compact GKS for compressible flow simulations, J. Comput. Phys. 549 (2026) 114612.

\bibitem{simple-WENO-1} J. Zhu, J.X. Qiu, A new fifth order finite difference WENO scheme for solving hyperbolic conservation laws, J. Comput. Phys. 318 (2016) 110-121.

\bibitem{simple-WENO-2} J. Zhu, J.X. Qiu, New finite volume weighted essentially non-oscillatory scheme on triangular meshes, SIAM J. Sci. Comput. 40 (2018) 903-928.

\end{thebibliography}
\end{document}